\documentclass[traditabstract]{aa} 
\usepackage{longtable}

\usepackage{graphicx}
\begin{document}
   \title{Kinematics of galaxies in Compact Groups\thanks{Based on observations collected at the European Southern Obervatory, La Silla, Chile.} \thanks{Figs. \ref{maps_hcg37d} to \ref{rot_cur3} are only available in electronic form.}}

   \subtitle{Studying the B-band Tully-Fisher relation}

   \author{S. Torres-Flores
          \inst{1,2},
	  C. Mendes de Oliveira\inst{1}, 
          P. Amram\inst{2},
	  H. Plana\inst{2,3},\\
	  B. Epinat\inst{4},
	  C. Carignan\inst{5,6},
	  C. Balkowski\inst{7}.
	  }

   \institute{Departamento de Astronomia, Instituto de Astronomia, Geof\'isica e Ci\^encias Atmosf\'ericas da USP, Rua do Mat\~ao 1226, Cidade Universit\'aria, 05508-090, S\~ao Paulo, Brazil
         \and
             Laboratoire d'Astrophysique de Marseille, OAMP, Universit\'e de Provence \& CNRS, 38 rue F. Joliot--Curie, 13388 Marseille, Cedex 13, France
         \and   
             Laborat\'orio de Astrof\'isica Te\'orica e Observacional, Universidade Estadual de Santa Cruz, Ilheus, Brazil
         \and
             Laboratoire d'Astrophysique de Toulouse-Tarbes, Universit\'e de Toulouse \& CNRS, 14 Avenue Edouard Belin, 31400 Toulouse, France 
	 \and
	     Laboratoire d'Astrophysique Exp\'erimentale (LAE), Observatoire du Mont M\'egantic, and D\'epartement de Physique, Universit\'e de Montr\'eal, C.P. 6128, Succ. Centre-Ville, Montr\'eal, QC, Canada H3C 3J7
         \and
	    Observatoire d'Astrophysique de l'Universit\'e de Ouagadougou (UFR/SEA), 03 BP 7021 Ouagadougou 03, Burkina Faso
         \and   
             GEPI, Observatoire de Paris, Paris University Denis Diderot \& CNRS, 5 place Jules Janssen, Meudon, France
             }

   \date{Received ; accepted }

 
  \abstract
{We obtained new Fabry-Perot data cubes and derived
velocity fields, monochromatic
and velocity dispersion maps for 28 galaxies in the Hickson compact groups
37, 40, 47, 49, 54, 56, 68, 79 and 93. We also derived
rotation curves for 9 of the studied galaxies, 6 of which are strongly 
asymmetric. 
Combining these new data with previously published 2D kinematic maps
of compact group galaxies, we investigate the differences between the
kinematic and morphological position angles for a sample of 46 galaxies.
We find that one third of the non-barred compact group galaxies have
position angle misalignments between the stellar and gaseous components.
This and the asymmetric rotation curves are clear signatures of kinematic
perturbations, probably due to interactions among compact group galaxies.
A comparison between the B-band Tully-Fisher relation for compact group galaxies and that for the GHASP field-galaxy sample shows that, despite 
the high fraction of compact group galaxies with asymmetric rotation
curves, these lie on the Tully-Fisher relation defined by galaxies in less
dense environments, although with more scatter. This is in agreement with
previous results, but now confirmed for a larger sample of 41 galaxies.
We confirm the tendency for compact group galaxies at the low-mass
end of the Tully-Fisher relation (HCG 49b, 89d, 96c, 96d and 100c)
to have either a magnitude that is too bright for its mass (suggesting
brightening by star formation) and/or a low maximum rotational velocity for
its luminosity (suggesting tidal stripping).  These galaxies are outside
the Tully Fisher relation, at the 1$\sigma$ level, even when the minimum
acceptable values of inclinations are used to compute their maximum
velocities.
The inclusion of such galaxies with v$<$100 km s$^{-1}$
in the determination of the zero point and slope of the compact group
B-band Tully-Fisher relation would strongly change the fit, making it
different from the relation for field galaxies, a fact that has to be 
kept in mind when studying scaling relations of interacting galaxies,
specially at high redshifts.}
{}{}{}{}
\authorrunning{S. Torres-Flores et al.}

   \maketitle
%

\section{Introduction}

Compact groups of galaxies are environments in which tidal encounters are
thought to be common. It is, therefore, expected,  that interactions have
stripped or disturbed such systems at some level. Mendes de Oliveira et
al. (2003) studied the Tully-Fisher (TF) relation for 25 compact group
galaxies and highlighted the importance of having 2D velocity fields, derived from the H$\alpha$ line, in the study of TF of interacting galaxies. They find similar
TF relations for compact group and field galaxies and stress that a fine
tuning of the kinematic parameters is needed in order to have meaningful
rotation curves for interacting galaxies. This is highly relevant for
the study of the evolution of the TF relation as a function of redshift,
given that at high redshifts the numbers of interacting galaxies raise
considerably. We are gathering a large dataset of Fabry-Perot velocity
maps for interacting galaxies at low redshifts which allows the
study of the effects of the environment on galaxy evolution in high density structures like compact groups of galaxies. With this goal in mind, we studied, in previous papers, the properties of the galaxies in these systems, pointing out interaction indicators for individual galaxies (e.g. Mendes de Oliveira et al. 1998); the evolution of the B-band TF relation (Mendes de Oliveira et al. 2003); the distribution of luminous and dark mass profiles of compact group galaxies (Plana et al. 2010).

The present study comes to complement the sample of compact group
galaxies with measured velocity maps presented by Plana et al. (1998, 2000, 2003), Mendes de Oliveira et al. (1998), Amram
et al. (2003, 2004, 2007) and Torres-Flores et al. (2009). We present, for the first time, velocity maps for galaxies in nine Hickson
compact groups (HCGs 37, 40, 47, 49, 54, 56, 68, 79 and 93.) and we revisit the
B-band Tully-Fisher relation for a sample of 41 Hickson compact group
galaxies, including a comparison with the GHASP sample of field galaxies,
for which 2D velocity fields, derived from the H$\alpha$ line, are also available (Epinat et al. 2008b).


\section{Observation and data reduction}



%
  
%

\subsection{The Sample}

We obtained new Fabry-Perot data for galaxies in the following Hickson
Compact Groups (HCG) HCG 37, HCG 40, HCG 47, HCG 49, HCG 54, HCG 56,
HCG 68, HCG 79 and HCG 93. This sample is composed by nearby groups from
the sample of Hickson et al. (1992), with radial systemic velocities
lower than 10500 km s$^{-1}$. In order to study the relation between
kinematic and morphological parameters, we have considered the galaxies
in this study for which it was possible to determine the kinematic parameters 
(11 galaxies of a total of 28). In addition we considered the systems studied in Torres-Flores et
al. (2009, 7 galaxies) and those of Mendes de Oliveira et al. (2003, 32 galaxies), for which kinematic and morphological measurements
could be made or were already available in the literature. In some cases, 
there was no
information about the kinematic position angle and inclination of the
galaxies (HCG 10a, 88c and 2c, 10a, 22c, 40e, 88c, respectively). In other cases, no error
bars for the position angle and inclinations were quoted (HCG 7c and 79d). All these
objects were excluded from our analysis. Finally, we compare kinematic and morphological position angles for 46 galaxies and kinematic and morphological inclinations for 43 galaxies.

In the case of the
B-band Tully-Fisher relation, we considered the galaxies of this study (9 galaxies with measured rotation curves of a total of 28),
7 galaxies from Torres-Flores et al. (2009) and 25 galaxies (of a total of 32 galaxies) studied by Mendes
de Oliveira et al. (2003). In the
case of HCG 79d, we did not derive its rotation curve and kinematic
parameters given that they have already been published in 
Mendes de Oliveira et al. (2003).

\subsection{Observations}

The observations were carried out using a Fabry-Perot instrument mounted either on the Canada France Hawaii telescope (MOSFP, Amram et al. 2003) or
on the European Southern Observatory 3.6m telescope (CIGALE, Amram et al. 1991), using either a CCD or
a photon counting system respectively. The observed
compact groups, exposure times and scanning wavelengths are listed in
Table \ref{table1}. A journal of the observations is given in Table
\ref{table2}. 

\begin{table}
\begin{minipage}[t]{\columnwidth}
\caption{Journal of observations}
\label{table1}
\centering
\renewcommand{\footnoterule}{}  
\begin{tabular}{cccc}
\hline \hline
HCG & Telescope \& & Exp. Time & Obj. Scan \\
  &   date of obsservation  & (hours)& Wavelength \AA \\
\hline
2ab & ESO Sep 2000  & 1.2 & 6657 \\
2c & ESO Sep 2000  & 0.9 & 6657  \\
7ad  & ESO Sep 2000  & 1.5 & 6655 \\
7bd  & ESO Sep 2000  & 0.9 & 6655  \\
22bc & ESO Sep 2000  & 1.1 & 6621 \\
37a & CFHT March 2000 & 2.1 & 6710 \\
37d & CFHT March 2000 & 2.1 & 6710 \\
40 & CFHT March 2000 & 1.9  & 6709 \\
47 & CFHT March 2000 & 2.3  & 6771 \\
49 & CFHT March 2000 & 2.3  & 6780 \\
54 & CFHT March 2000 & 1.9  & 6594  \\
56 & CFHT March 2000 & 2.3  & 6747  \\
68ab & CFHT March 2000 & 1.9  & 6613 \\
68c & CFHT March 2000 & 1.9  & 6616 \\
79 & CFHT Aug 1996 & 1.2 & 6661  \\
93ac & CFHT Aug 1996 & 1.2 &  6675 \\
93b & CFHT Aug 1996 & 1.2  & 6665  \\
\hline
\end{tabular}
\end{minipage}
\end{table}

\begin{table*}
\begin{minipage}[t]{\textwidth}
\caption{Instrumental setup}
\label{table2}
\centering
\renewcommand{\footnoterule}{}  
\begin{tabular}{llll}
\hline \hline
\multicolumn{1}{c}{Parameters} & \multicolumn{3}{c}{Values}\\
\hline
\hspace{-0.2cm}HCG observation: \\
Telescope & CFHT 3.6m  & CFHT 3.6m & ESO 3.6m  \\
Equipment  &  MOS/FP &  MOS/FP  & CIGALE  \\
Date   & 1996 Aug  & 2000 Mar &  2000 Sep  \\
\hspace{-0.2cm}Calibration: \\
Neon light   & $\lambda$ 6598.95 \AA  & $\lambda$ 6598.95 \AA & $\lambda$ 6598.95 \AA   \\
Interferometer order at H$\alpha$  & 1162  & 1162 & 793 \\
Free spectral range at H$\alpha$ (km s$^{-1}$) & 265  & 258 &  378 \\
\hspace{-0.2cm}Sampling: \\
Number of scanning steps   & 24  & 28 & 32 \\
Sampling steps \AA (km s$^{-1}$) & 0.24 (11.0)  & 0.21 (9.2) & 0.26 (11.8) \\
\hspace{-0.2cm}Detector:
    &  CCD & CCD &  IPCS \\
\hline
\end{tabular}
\end{minipage}
\end{table*}

\subsection{Data Reduction}

The reduction method we used was the same as the 
one applied to the GHASP sample in Epinat et al. (2008a,b) who used the reduction package developed by Daigle et al. (2006b). The output of this package consists of 
monochromatic, continuum, dispersion and radial velocity maps. This
procedure includes several improvements with respect to the data reduction
and analysis performed to the data presented in Plana et al. (2003)
and Amram et al. (2003), listed in the following.  First, the present
procedure uses an adaptive spatial binning to the data based on the 2D
Voronoi tessellation method (see Cappellari \& Copin 2003) applied to the 3D data cubes, 
optimizing the determination of the spatial resolution. In this process, the signal to
noise ratio (SNR) of a bin is computed every time a new pixel is added
to the bin. When a required SNR (previously defined by
the user) is reached, no more pixels are added to the analyzed bin.
Second, the velocity field is automatically cleaned, i.e. 
all the residual features that are not linked with emission lines 
in the objects in the groups are removed. Third, we compute the
astrometric correction comparing stars in DSS images with the continuum
image generated from the cube using KOORDS task in KARMA software. In the present work, the SNR was estimated using the square
root of the flux. When the addition of pixels gave a SNR of 6 or 8 (for photon counting system or CCD, respectively), no more pixels were added to the bin. We have used Gaussians to achieve spectral smoothing. 
Wavelength calibrations were obtained by scanning the narrow Ne 6599 $\rm{\AA}$ line under the same conditions as the observations. The OH subtraction was performed by estimating the sky using the medium spectrum of the data cube.

\section{Data analysis}

\subsection{Morphological and kinematic parameters}

Inclinations, position angles and galactic centers can be determined either from the morphology (using broad-band imaging) or from the kinematics (using the velocity field). 

The morphological inclinations were computed using the axial ratio ${\rm cos}(b/a)=i$, where b and a are the optical diameters at 
the 25 mag arc$^{-2}$ isophote (Hickson, 1993). Due to the distorted morphologies of the compact group galaxies, we avoid using 
the morphological type of the galaxy for computing the inclinations. The morphological centers are estimated using either the 
center of best ellipse fitting or the flux peak in the continuum image. Due to the asymmetries in the morphology, we have used 
the second method. The morphological PAs are estimated using best ellipse fitting on the continuum image at the isophote of 25 mag arcsec$^{-2}$ (R$_{25}$).

Kinematic centers, inclinations, PAs and systemic
velocities were estimated from the velocity fields using the procedure
developed in Epinat et al. (2008a). We list these parameters in Table
\ref{table3}. The procedure, in three steps, we used to determine these parameters is the following. In a first attempt, all the kinematic parameters were
estimated automatically. In a second one, we fixed the center using the
morphological center and we allowed
the inclination and position angle of the galaxies to be automatically
measured in the computation process (from the velocity fields). In a final attempt, we left
only the position angle free, to be obtained from the velocity map, and we fixed the center
and the inclination, using the morphological values. From these three steps, we concluded 
that using the morphological inclination and center and the kinematic PA we obtain 
more symmetrical rotation curves than in other cases.

\begin{figure}[t!]
\hspace{-0.70cm}
\includegraphics[scale=0.75]{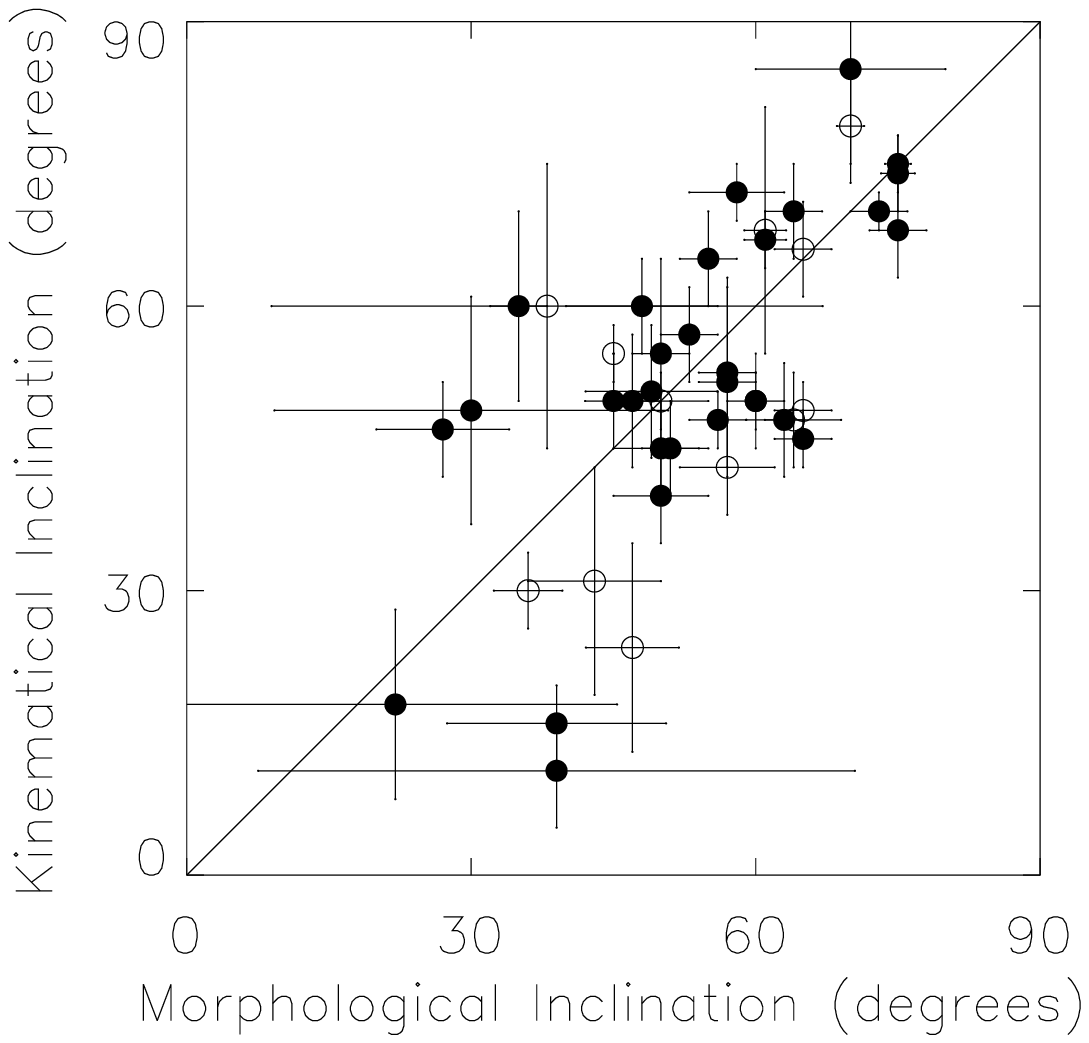}\\

\hspace{-0.60cm}
\includegraphics[scale=0.55]{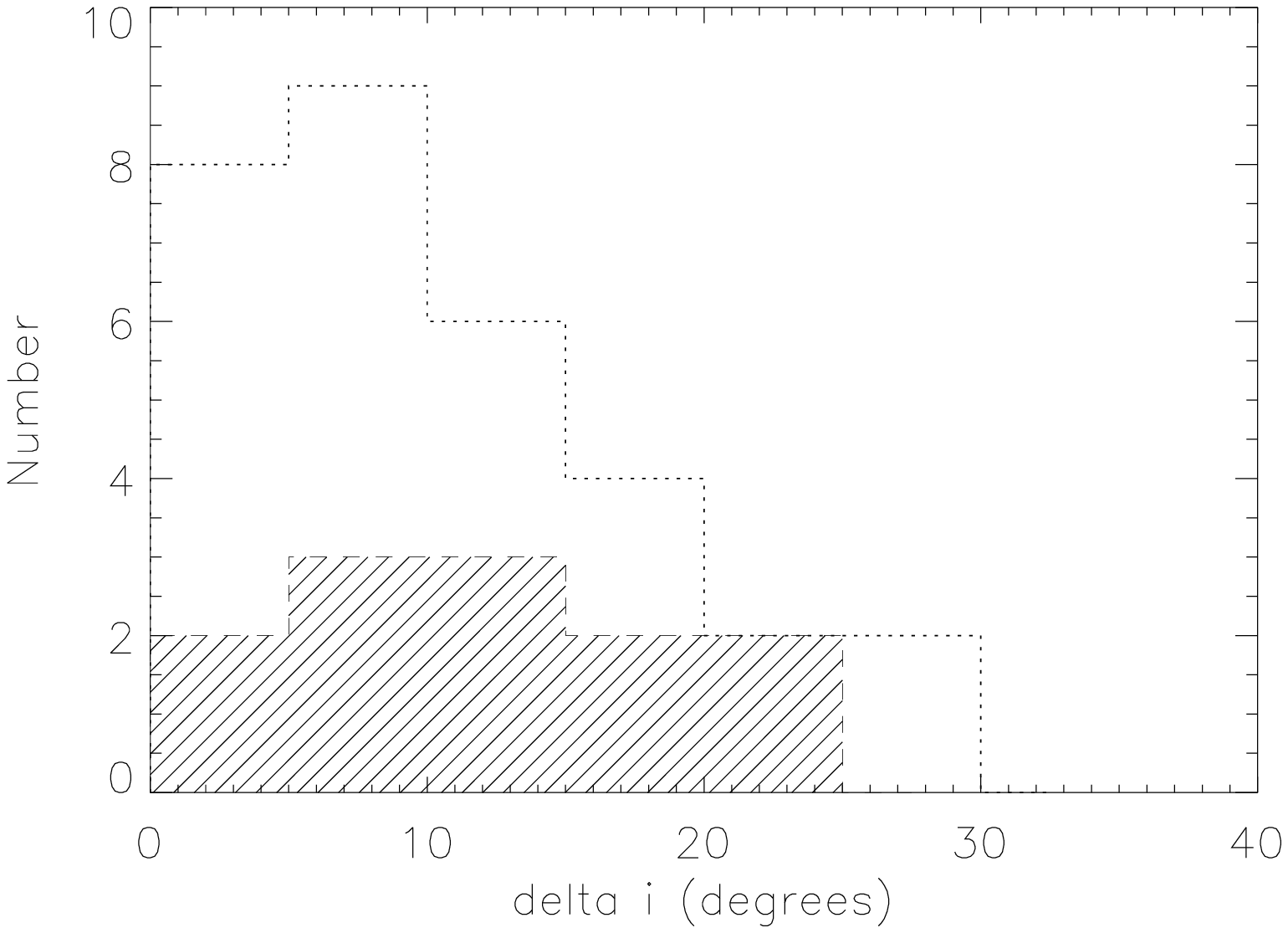}
\caption{Top panel: Kinematic versus morphological inclination for the 43 HCG galaxies with measured inclinations. 
Open and filled circles correspond to the barred and non-barred galaxies respectively. Bottom panel: Histogram of the differences between the kinematically and morphologically-derived inclinations. 
Non-barred galaxies are represented by a dotted line (open 
histogram). Dashed lines represents barred galaxies (filled histogram).}
\label{incli}
\end{figure}

\begin{figure}[t!]
\hspace{-0.70cm}
\includegraphics[scale=0.75]{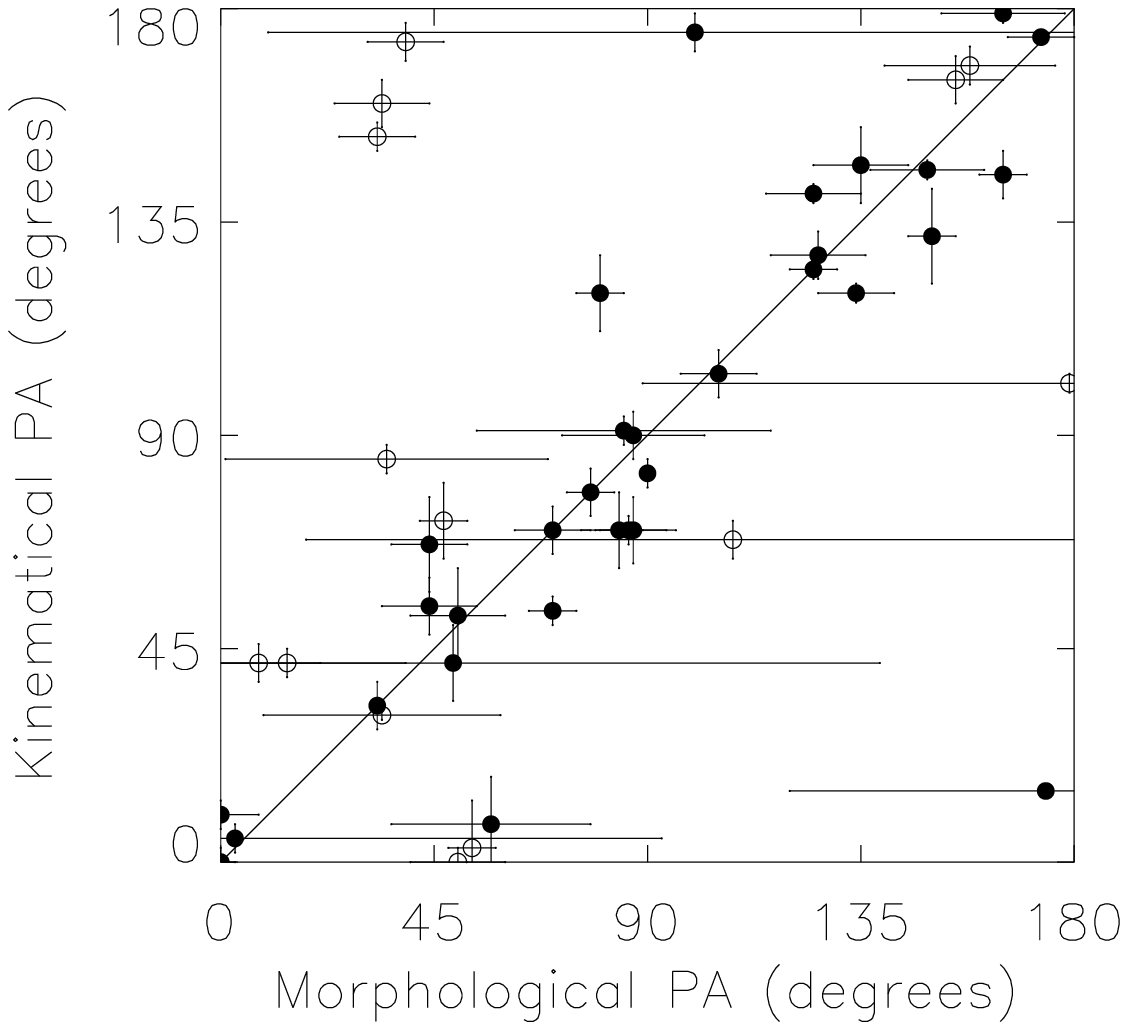}\\

\hspace{-0.60cm}
\includegraphics[scale=0.55]{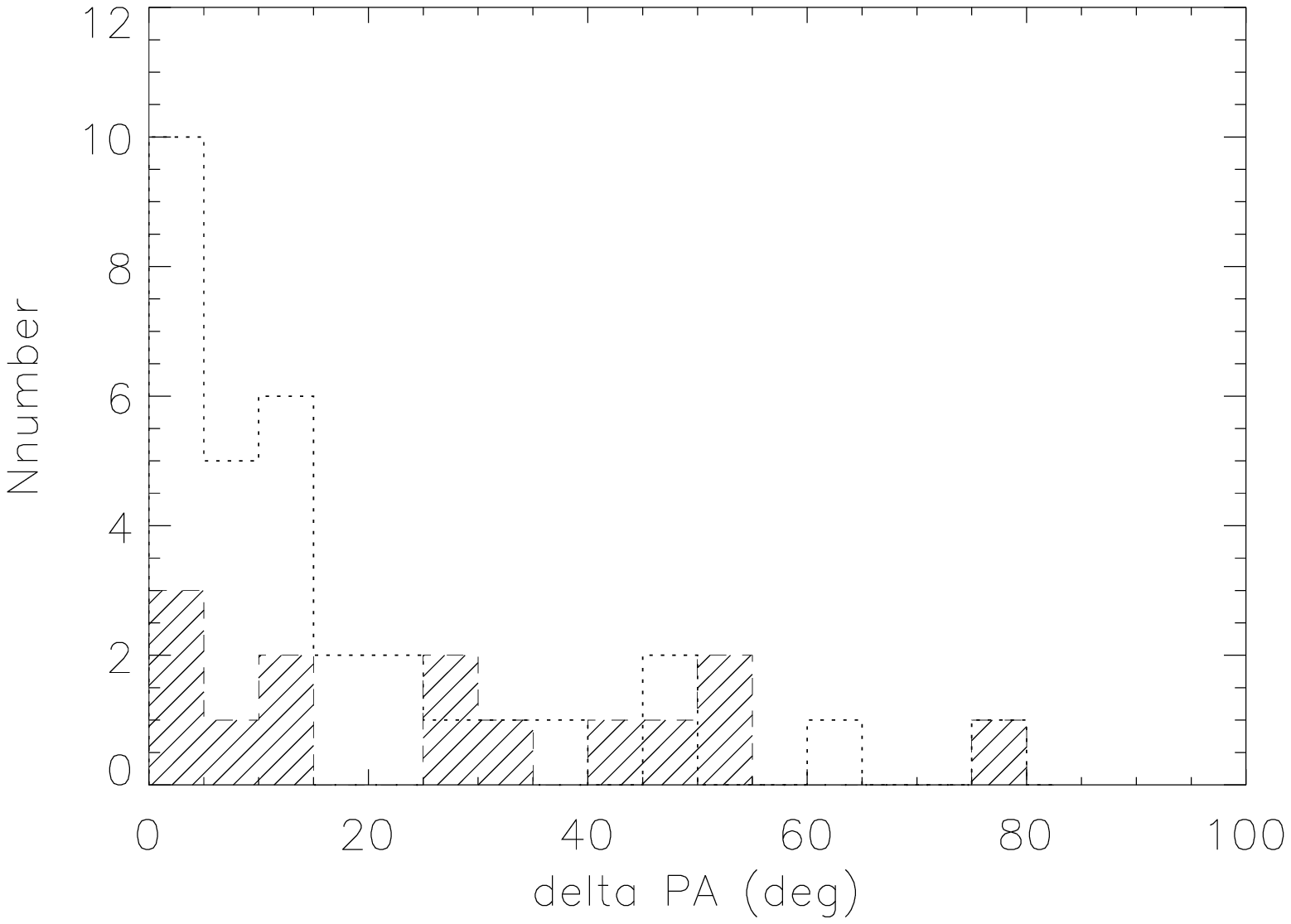}
\caption{Top panel: Kinematic versus morphological position angle for the 46 HCG galaxies with measured position angles. 
Open and filled circles correspond to the barred and non-barred galaxies respectively. Bottom panel: Histogram of the differences between the kinematically and morphologically-derived position angles. 
Non-barred galaxies are represented by a dotted line (open histogram). 
Dashed lines represent barred galaxies (filled histogram).}
\label{pa}
\end{figure}

\begin{figure}[t!]
\hspace{-0.93cm}
\includegraphics[scale=0.57]{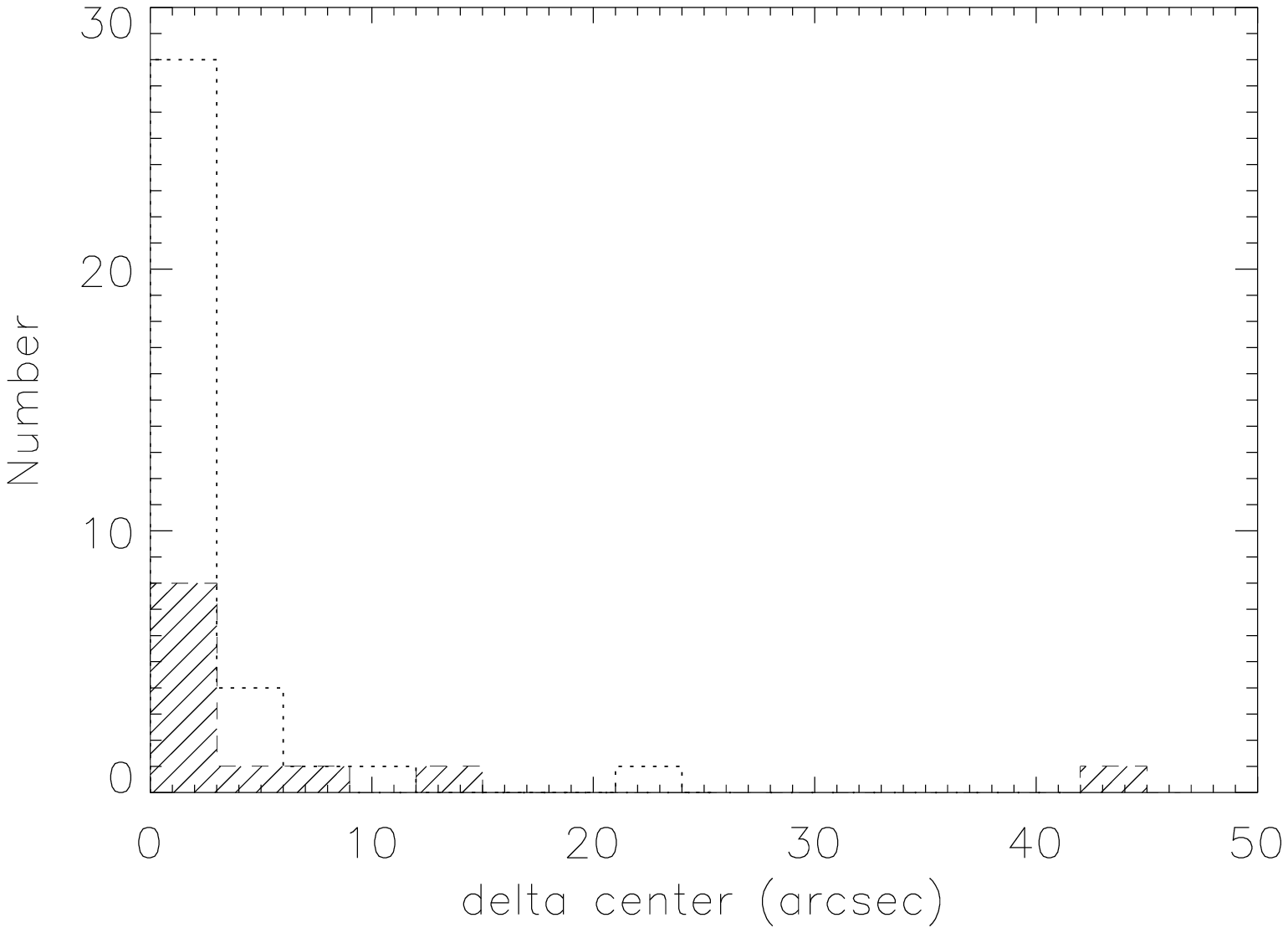}
\caption{Histogram of the differences between 
kinematically- and morphologically-derived centers for 48 HCG galaxies with measured centers. Non-barred galaxies are represented by a dotted line (open histogram). Dashed lines represents barred galaxies (filled histogram).}
\label{histo_cent}
\end{figure}

In Figs. \ref{incli} to \ref{histo_cent} we plot comparisons of
the kinematically- and morphologically-derived position angles, 
inclination and centers of the compact group galaxies. We included HCG 91c1 and 91c2 as two differents 
points in the diagrams, as studied by Mendes de Oliveira et al. (2003). In
Fig. \ref{histo_cent} we did not plot galaxies 
HCG 47a and 47d, given that the
automatic routine was not able to find the correct kinematic centers for
these galaxies.

Fig. \ref{incli} (top panel) shows that there is a relatively good agreement between the kinematically and morphologically derived inclinations, except in a few cases. Nevertheless, due to the fact that the velocity fields are not regular enough, the inclination can not be safely determined from the kinematics. Thus, in the following analysis, we preferred to use the morphological inclination deduced from the optical images, which are usually more symmetric.

In Fig. \ref{incli} (bottom panel), the histogram shows 11 galaxies having a 
difference in inclination between the 
morphologically and kinematically derived values (delta i) 
of more than 15 degrees. 
These are HCG 2b, 10c, 16d, 19b, 37d, 47a, 47d, 49c, 89b and 89c.

The determination of the position angle for the GHASP galaxies
(Epinat et al. 2008a) was obtained through an automatic procedure using
the velocity fields of the galaxies, which worked well for isolated galaxies.
We made similar tests for the compact group sample and
concluded that the automatic estimates using the velocity fields are
robust. We checked the quality of the kinematic position angle determination by visual inspection.

Fig. \ref{pa} (top panel) shows that optical PAs have large error bars. On the other hand, kinematic PAs present error bars smaller than 10 degrees. Due to this result, we used the kinematic PA in this study.

In Fig. \ref{pa} (bottom panel) we show the histogram of the differences between the kinematically- and photometrically-determined position angles. In this figure, non-barred galaxies are represented by dotted lines (open histogram). Barred galaxies are shown as dashed lines (filled histograms). Most of the non-barred galaxies have delta PA lower than 20 degrees while the barred galaxies do not show a trend in the plot. The delta PA for barred galaxies span from 0 to 80 degrees with no peak in the distribution. Peirani et al. (2009) found that given specific initial conditions, an interaction between two galaxies having a mass ratio of 1:3 could produce a bar. In view of this result, we can not exclude as interacting objects the barred galaxies that present a misalignment between the optical and kinematic position angles lower than 20 degrees. In the case of galaxies having delta PA higher than 20 degrees, it could be easily associated with galaxy-galaxy interactions.

In Fig. \ref{histo_cent}, the histogram shows the difference (in arcsec) between the kinematic and morphological center for each galaxy, when all parameters (inclination, position angle and center) are not fixed, 
i.e., are left for the program to fit.  These values are listed in Table
\ref{table3}. For HCG 2a, HCG 7d and HCG 49c there is a large shift
between the kinematic and morphologic center, suggesting the inadequacy
of the automatic fit. It can be noted in Fig. \ref{histo_cent},
that these three galaxies have delta center larger than 10 arcsec. We exclude of this analysis HCG 47a and 47d, given that it was not possible to derive the kinematic center of these objects.

As expected, the most symmetrical rotation curves were obtained by fixing the
center and inclination (from the morphology) and letting the position angle be determined automatically.

No kinematic analysis was performed for galaxies in HCG 54, due to their anomalous nature (Verdes-Montenegro et al. 2002). 

\subsection{Maximum rotational velocities}

Rotation curves and error bars were computed using the 
parameters derived in \S 3.1 and following the same procedure as
Epinat et al. (2008a). In order to study the Tully-Fisher relation for
the compact group galaxies, we estimate the maximum rotational velocity
from each rotation curve. To enlarge the statistics, we also include all galaxies
studied by Mendes de Oliveira et al. (2003) for which maximum rotational velocities were defined as the maximum
values for the average velocities for the approaching and receding sides, in a
similar way as was defined by Epinat et al. (2008a) for edge-on galaxies. 

For several galaxies (HCG 2b, 7a,
7d, 37d, 40c, 40e, 49a, 49b, 49c, NGC 92) the rotation curves do not
reach the optical radius (R$_{25}$). In these cases we fit an arctan
function ($V=V_{0}\times(2/\rm{\pi})\times \rm{arctan}(\it{r/r_{t}})$, Courteau 1997) in order to
estimate the velocity at R$_{25}$, which will be used as the maximum
rotational velocity. Due to the inner shape of the rotation curve, in two cases (HCG 2b and HCG 40e) this fit
gives us extremely overestimated (and clearly erroneous) maximum rotational 
velocities (see Appendix B.). In those cases, we take the average between the velocity at R$_{25}$
from the arctan fit and the maximum velocity measured in the rotation curve. 
All these values were checked by eye on the rotation curves. At the end, 41 galaxies are available to study the Tully-Fisher relation in compact groups. 
Table \ref{table4} lists the observed and modeled rotational velocities for the unpublished galaxies.

\subsection{Comparison with field galaxies}

In order to investigate if the Tully-Fisher relation of compact group galaxies
is similar to that for field galaxies, we used as a control sample the
galaxies of the GHASP survey.  Epinat et al. (2008a,b) found that the
galaxies in this sample lie on the B-band Tully-Fisher relation,
in good agreement with previous determinations of TF for field galaxies
(Tully \& Pierce 2000). We exclude from their sample all the galaxies
with radial systemic velocities lower than 3000 km s$^{-1}$, for
which no other individual measurements of distances were available
(the references are indicated in Epinat et al. 2008b). We also removed
from the list galaxies with inclinations lower than 25 degrees,
because of the large uncertainties in the velocities. Finally, we did not use
galaxies for which the maximum velocities were not reached
(Epinat et al. 2008b). This left us with 93 galaxies in the GHASP sample.

\begin{table*}
\begin{minipage}[t]{\textwidth}
\caption{Kinematic properties of the sample}
\label{table3}
\centering
\renewcommand{\footnoterule}{}  
\begin{tabular}{cccccccccccc}
\hline \hline
(1)\footnotetext{Column (1): Group number. Column (2): Galaxy member of the group. Column (3): PA deduced from our velocity field. Column (4): PA taken from Mendes de Oliveira (1992) for HCG. This was determinated for semimajor axis lengths of 2h$^{-1}$kpc for all Hickson group galaxies with z$\leq$0.05. Column (5): Optical PA taken from HyperLeda. PA is the position angle of the major axis of the isophote 25 mag arcsec$^{-2}$ in the B-band. Column (6): Inclination deduced from the analysis of our velocity field. Column (7): Morphological inclination deduced from the axis ratio of the isophote 25 mag arcsec$^{-2}$ in the B-band for galaxies ($\rm{cos}\it{(b/a)}=\it{i}$). For NGC 92 given values correspond to HyperLeda/NED. Column (8): Morphological inclination taken from HyperLeda. Column (9): Systemic velocity taken from
Hickson (1993). Column (10): Systemic velocity deduced
from our velocity field analysis. Column (11): Radial velocity corrected by the Virgo infall from HyperLeda. Column (12): Difference in arcseconds between the morphologial and kinematic centers.}&(2)&(3)&(4)&(5)&(6)&(7)&(8)&(9)&(10)&(11)&(12)\\ 
HCG & Galaxy & PA$_{\rm{kin}}$ & PA$_{\rm{morph}}$ & PA$_{\rm{morph}}$
&i$_{\rm{kin}}$ &i$_{\rm{morph}}$ & i$_{\rm{morph}}$ &V$_{\rm{sys HICK}}$ & V$_{\rm{sys FP}}$&
vvir & $\delta$ c$_{\rm{kin-morph}}$\\ & & deg & deg & deg & deg & deg &
deg &km s$^{-1}$ & km s$^{-1}$& km s$^{-1}$ & arcsec\\
\hline
2
&a &  42$\pm$4 & 3 & 8$\pm$13   & 68$\pm$13 & 61$\pm$2 & 65 & 4326 & 4347& 4375  &42.2\\
&b &  5$\pm$3 & 28  & 3$\pm$90   & 49$\pm$12 &  30$\pm$21  & 43 & 4366  & 4357&  4365 &1.3\\
&c &  168$\pm$4 & 132 & 158$\pm$18 & ... &   57$\pm$3 & 64 & 4235 & 4266& 4284  &2.4 \\
7
&a &  179$\pm$2 & 153 & 165$\pm$13  & 67$\pm$3 & 61$\pm$2  &  76 & 4210& 4198& 4178   &0.6\\
&d &  85$\pm$3 & 58  & 35$\pm$34   & 31$\pm$12  & 43$\pm$7 &   44 & 4116&4104&  4081 &12.3\\
22
&c &  101$\pm$2 & 70  & 179$\pm$90 & ...  &  25$\pm$14 & 90  & 2728 & 2572  & 2444 &6.8\\
37
&d   &  68$\pm$4 & 58  & 108$\pm$90 &  60$\pm$15 & 38$\pm$29 &36 & 6207 & 6308& 6236 &0.8\\
40
&c   &  120$\pm$2 & 125 &  134$\pm$8 & 74$\pm$4  &  75$\pm$2 &90  & 6890 & 6888& 6341 &1.6\\
&e   &  8$\pm$10 & 72  &  57$\pm$21 &  ... & 70$\pm$6  & 80 & 6625 & ...&  6383  &0.0\\
47
&a   & 42$\pm$3 & 42 & 14$\pm$25  &  24$\pm$11 & 47$\pm$5 &  67 &9581 & 9507 & 9622   &...\\
&d   & 15$\pm$1 & 165& 174$\pm$54 & 16$\pm$4  & 39$\pm$12 &  32 & 9471 & 9325& 9533&...\\
49
&a   & 42$\pm$8 & 57 & 49$\pm$90 & 18$\pm$10  & 22$\pm$23 &  43 &9939 & 9812& 10196 &9.1\\
&b   & 91$\pm$3  & 72 & 85$\pm$31 &  51$\pm$7 & 49$\pm$7  & 59 & 9930& 9787& 10190&4.5\\
&c   & 175$\pm$4 & 98 & 100$\pm$90 &  11$\pm$6 & 39$\pm$31  & 45 & 9926& 9784& 10185 &23.7\\
56
&a   & 174$\pm$1 & 172 & 173$\pm$7  & 75$\pm$3  & 75$\pm$1  &  90  & 8245& 8374&8455  &1.6\\
68
&c   & 31$\pm$1 & 119 & 34$\pm$25   & 30$\pm$4  & 36$\pm$4 &  54 & 2313 & 2515& 2550 &1.0\\
93
&b   & 173$\pm$4 & 156 & 39$\pm$8  &  79$\pm$6 & 70$\pm$1 &  68 & 4672 & 4625& 4776 &1.3\\
&NGC 92 & 146$\pm$2 & 149 & 149$\pm$12 & 48$\pm$6 & 63/62$\pm$2 & 70 & 3219 & 3412 & 3132 &0.7\\
\hline
\end{tabular}
\end{minipage}
\end{table*}

\subsection{Magnitudes}

The absolute magnitudes were obtained using $M=B_{\rm{cor}}-5\times(\rm{log}(\it{vvir}_{\rm{group}}\rm{/75))-25}$, 
where $\it{vvir}_{\rm{group}}$ corresponds to the heliocentric radial velocity of the group
to which the galaxy belongs, 
corrected for local group infall onto Virgo, from HyperLeda (Paturel et al. 2003)\footnote{http://leda.univ-lyon1.fr}. B-band magnitudes (B$_{\rm{cor}}$) were taken from B$_{\rm{T}}$, from Hickson et al. (1989). We note that for 44 $\%$ of the galaxies used in our analysis the magnitude errors are lower than 0.1, for 93$\%$ it is lower than 0.2 mag. For HCG 7c, 22c and 89a the errors are larger, therefore, we checked their magnitudes in the Hyperleda database.  The magnitude for HCG 22c is in agreement with the value reported by Hyperleda, however, for 7c and 89a the difference in magnitudes between Hickson et al. (1989) and Hyperleda are large (0.7 and 0.5 mag, respectively).  These differences in magnitude are consistent with the errors of 0.7 and 0.5 mag given in Hickson et al. (1989), respectively for both galaxies. The B$_{\rm{T}}$ magnitudes were then corrected by Galactic extinction (Schlegel et al. 1998) and by the extinction due to the inclination (Bottinelli et al. 1995). Values for the galactic extinction and
extinction due to the inclination were taken directly from HyperLeda.
Details on the B-band magnitudes for the GHASP sample are given in Epinat et al. (2008a,b). Table \ref{table4} lists the corrected B-band absolute magnitudes for all HCG galaxies.

\begin{table*}
\begin{minipage}[t]{\textwidth}
\caption{Rotational velocities of the sample}
\label{table4}
\centering
\renewcommand{\footnoterule}{}  
\begin{tabular}{ccccccccccc}
\hline \hline
& & & &\multicolumn{4}{c}{Observed} &\multicolumn{1}{c}{Model} \\
\hline
(1)\footnotetext{Column (1): Group number. Column (2): Galaxy member of the group. Column (3): Morphological type taken from Hickson (1993). Column (4): B-band absolute magnitude. Column (5): Isophotal radius at the limiting surface brightness of 25 B mag arcsec$^{-2}$, Column (6): Observed rotational velocity at R$_{25}$. Column (7): Maximum radius reached by the rotation curve. Column (8): Observed maximum rotational velocity. Column (9): Rotational velocity at R$_{25}$ derived from the arctan model. Column (10): Rotational velocity used in the Tully-Fisher relation. Column (11): Quality flag on V$_{\rm{max}}$ (1: reached; 2: probably reached; 3: probably did not reach; 4: did not reach.)}&(2)&(3)&(4)&(5)&(6)&(7)&(8)&(9)&(10)&(11)\\
HCG& Galaxy & Morph.& M${\rm_{B}}$ & R$_{25}$ & V$_{\rm{R(25)}}^{\rm{RC}}$ & R$_{\rm{max}}$ & V$_{\rm{max}}^{\rm{RC}}$ & V$_{\rm{R(25)}}^{\rm{model}}$ & V$_{\rm{max}}^{\rm{TF}}$ &V$_{\rm{max}}$ \\  
&&Type&mag&(arcsec/kpc)&km s$^{-1}$& (arcsec/kpc) & km s$^{-1}$ & km s$^{-1}$ & km s$^{-1}$ & \\ 
\hline
2   	    
&a & SBd & -20.75  &41/11.4&212&74/21.0& 264& 197 &  264$\pm$39 & 1 \\
&b & cI &  -19.39  &21/6.0 &...&7/2.0& 122& 270 &  196$\pm$74 & 4 \\
&c & SBC &  -19.87 &33/9.1 &118&40/11.2& 122& 106 &  122$\pm$13 & 1 \\
7  		               
&a & Sb & -20.69   &57/15.3 &...&47/12.7& 226& 197 &   226$\pm$27 & 2 \\
&d & SBC & -18.66  &38/10.1&...&16/4.3& 78 &  85 &  85$\pm$4 & 2 \\
22 		    	 
&c & SBcd & -19.48 &52/8.3&143&52/8.3& 142& 142 &   142$\pm$9 & 1 \\  
37 		    	 
&d & SBdm & -18.65 &12/4.8&...&5/2.0& 63 & 76  &   76$\pm$7  & 4  \\
40 		    	 
&c & Sbc & -19.91  &37/15.1&...&24/9.9& 197& 203 &  203$\pm$3  &  2 \\
&e & Sc  & -18.02  &18/7.3&...&7/2.9& 146&  226&  186$\pm$40  & 4 \\
47 		    	 
&a & SBb & -21.23  &...&...&...& ...& ... &  ...& ... \\
&d & Sd  & -19.33  &...&...&...& ...& ... &  ...& ... \\
49 		    	 
&a & Scd & -19.76  &14/9.4&...&10/6.4&126 & 182  &  154$\pm$28  & 4 \\
&b & Sd & -19.47   &13/8.7&...&4/2.5&69  & 58  &  58$\pm$6  & 3 \\
&c & Im & -18.36   &10/6.6&...&7/4.3&123 & 133 &   133$\pm$5 &4  \\
56 		    	 
&a &  Sc  &-20.52  &28/15.2&200&22/12.0&205 & 228 &  205$\pm$14 & 1 \\
68 		    	 
&c &  SBbc & -20.83&77/12.6&244&115/18.9&244 & 234 &  244$\pm$8 & 1 \\
93 		    	 
&b & SBd & -21.44  &57/17.7&191&28/8.5&235 & 220 &  235$\pm$22 & 2 \\
&NGC 92 &  Sa & -19.55&49/10.0 & ...&7/1.5&207 &  229  & 229$\pm$11 & 4 \\
\hline
\end{tabular}
\end{minipage}
\end{table*}

\subsection{Analysis of individual galaxies}

In the three appendices (A, B and C) to this paper, we present the optical B-band images from DSS, velocity fields, 
monochromatic images, velocity dispersion maps (Figs. \ref{maps_hcg37d} to \ref{maps_hcg93b}) and rotation curves 
(Figs. \ref{rot_cur1}, \ref{rot_cur2}, \ref{rot_cur3} and appendix C for rotation curves tables) for 11 galaxies in 
eight Hickson compact groups. Velocity dispersion maps presented are not corrected for instrumental nor thermal 
broadening. We analyze each individual galaxy using these new data together with literature data as listed in the 
following. We list in Table \ref{table3} the kinematic parameters obtained from the velocity fields (columns 3 and 8), 
the systemic velocities taken from the literature (column 9 and 11) and the systemic velocities resulting from our 
analysis (column 10). In Table \ref{table4} we list the different rotational velocities that we obtained, from the 
observations and from the arctan model (Courteau 1997). In column 6, we show the rotational velocties at R$_{25}$. 
Only 38$\%$ of the galaxies listed in Table \ref{table4} reaches R$_{25}$ (we noted that no rotation curve was computed for HCG 47a and 47d). Column 8 shows the observed maximum 
rotational velocity for each galaxy. Column 9 gives the modeled velocity at R$_{25}$. Column 10 corresponds to the 
rotational velocity used in the Tully-Fisher relation. We noted that for 39$\%$ of the sample V$_{\rm{max}}^{\rm{RC}}$=V$_{\rm{max}}^{\rm{TF}}$. 
Finally, in column 11 we flag the galaxies that reached, probably reached, probably did not reach or did not reach the maximum 
rotational velocity. We found that 38$\%$ of the galaxies in the sample did not reach their maximum rotational velocities.

\section{Results and Discussion}

\subsection{The Tully-Fisher relation}

\begin{figure}[ht!]
\hspace{-0.6cm}
\includegraphics[scale=0.55]{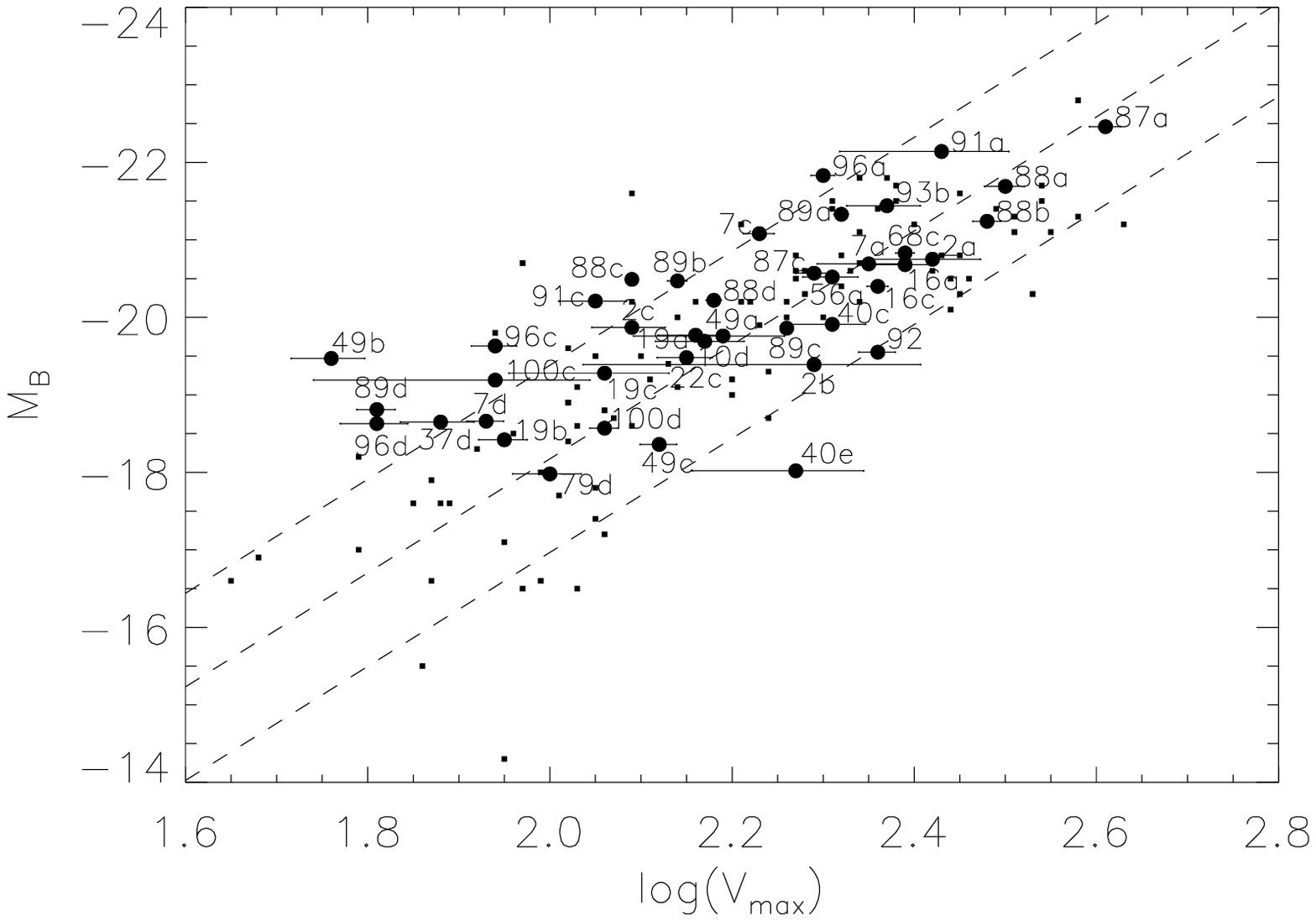}
\caption{B-band Tully--Fisher relation for galaxies in compact groups
(Mendes de Oliveira et al. 2003 and this work, filled circles) and a field
galaxy sample (Epinat et al. 2008b, small squares). For each rotation curve,
we derived the PA automatically, from the velocity field, 
using IDL routines. We fixed the center
using the morphological center and we fixed the inclination using the
morphological inclination.}
\label{TF_HCG_rc_ben_101}
\end{figure}

In Fig. \ref{TF_HCG_rc_ben_101} we show the B-band Tully-Fisher
relation for the HCG galaxies (filled circles, with numbers) and the GHASP sample (small
squares). Maximum rotational velocities were derived as described in \S 3.2. The dashed line in Fig. \ref{TF_HCG_rc_ben_101}
(equation 1) represents the best-fit least squares bisectors on the GHASP
data and 1$\sigma$.

\begin{equation}
M_{B}=(-3.47\pm1.21)-(7.35\pm0.53)\times[\rm{log}(\it{V}_{\rm{max}})]
\label{eq1}
\end{equation}

The slope of this fit is in agreement with the value obtained by Epinat
et al. (2008b) for the complete GHASP sample. Galaxies having
radial velocities lower than 3000 km s$^{-1}$ (and no independent distance
measurement) were removed from our analysis. The Tully-Fisher relation 
derived here for the GHASP sample is
in agreement with the results from Tully \& Pierce (2000), who found a
slope of -7.27. In order to compare our results with other ones 
from the literature, an inverse least-square fit was applied to the GHASP
sample. We found a slope of -9.11$\pm$0.85. Kannappan et al. (2002) and
Verheijen (2001) found slopes of -10.09$\pm$0.39 (using an unweighted
inverse fit) and -9.0$\pm$0.4 (RC/FD sample, excluding NGC 3992),
respectively. These results are also in agreement with the ones found in this
paper.

Figure \ref{TF_HCG_rc_ben_101} does not exhibit large differences in the
TF of HCG galaxies and the GHASP sample. Only one galaxy is off by more
than 2$\sigma$ from the TF relation (HCG 49b). We note that only one galaxy (HCG 40e) is more than 1$\sigma$ lower that the Tully-Fisher relation. Several cases are possible: it could have a high rotational velocity for its magnitude or a faint magnitude for its velocity or both. In the first case, the high maximum velocity we computed maybe is overestimated. Indeed, the maximum rotational velocity is observed only at $\sim$R$_{25}$/3 (146 km s$^{-1}$) and we extrapolated to 186 km s$^{-1}$ at R$_{25}$ (see section 3.2). If the actual velocity is 146 km s$^{-1}$ this galaxy lie on the 1$\sigma$ TF relation. On the other hand, if the maximum rotational velocity is 186 km s$^{-1}$ no inclination could push the galaxy on the 1$\sigma$ relation. In the second case, the problem is linked to the extinction. We can not rule out the possibility that the extinction of this galaxy may have been underestimated given the difficulty in making such measurements in an edge-on galaxy. That underestimated value would conspire to push the galaxy out of the TF relation. The galaxy has a B-band internal extinction of 0.41 (determined using Bottinelli et al. 1995).  To lie on the 1$\sigma$ relation, this galaxy should have and internal extinction of $\sim$1.3 mag, which is a very unlikely value. In conclusion, it is most likely that the maximum rotational velocity was overestimated for HCG 40e, due to the extrapolation in the rotation curve.

In the low mass regime, galaxy HCG 49b appears to have lower rotational velocity than the expected value from the TF relation or
alternatively it appears to have a higher B luminosity than that expected
from its observed maximum rotational velocity. This galaxy should have
a velocity of 151 km s$^{-1}$ to lie on the TF relation. An inclination
of 17 degrees could give us that rotational velocity. However, this
value is too low considering the kinematic and morphological inclinations
given in Table \ref{table3}. This effect was also noted by Mendes de Oliveira et al. (2003), for other galaxies at the low-mass end of the relation, e.g. HCG 96d and HCG 89d.

\subsection{Outliers from the Tully-Fisher relation}

Compact groups of galaxies are environments where interactions
between galaxies take place frequently. This is confirmed by a number of
studies which showed distorted kinematics and morphologies of compact
group galaxies (e.g. Presotto et al. 2009, Amram et al. 2007, Coziol \& Plauchu-Frayn 2007, Amram et al. 2003, Plana et al. 2003,
Plana et al. 1998 and Mendes de Oliveira et al. 1998),
also by studies of groups containing HI tidal debris (Verdes-Montenegro et al. 2001) and newly formed star forming
systems outside the luminous member galaxies (e.g. de Mello et al. 2008) as well as studies
of diffuse light presumably originated from stripped material from the
outskirts of galaxies in compact groups (e.g. da Rocha et al. 2008).
From these studies, it is clear that galaxy interactions represent an
effective mechanism to remove some of the stars and neutral gas from
the member galaxies. However, another important question arises: can interactions remove some of the dark halo of the compact
groups galaxies? Mendes de Oliveira et al. (2003) found a few low-mass
galaxies that had a lower maximum velocity than that expected from the
TF relation. In this paper we find one more galaxy, HCG 49b. Given that these studies focus on the the B-band Tully-Fisher relation, we can not
tell if these galaxies have been brightened by star formation or if they
have truncated profiles. In the latter case, if HCG 49b sometime 
in the past lied on the TF relation
(defined by the best fit to the GHASP data), this object would have had
to lose 85$\%$ of its possible initial mass to move to
its observed position on the TF relation. K-band Tully-Fisher relation and mass profile decomposition into luminous and halo components are necessary to address this issue (Torres-Flores et al. 2010c, 2010d).

In order to quantify the possible similarities/differences between the HCG and the GHASP samples, we apply a least square bisector fit to the HCG sample. In this case we obtain: $M_{\rm{B}}=(-8.40\pm1.20)-(5.30\pm0.53)\times[\rm{log}(\it{V}\rm{_{max}})]$. From these values and inspecting Fig. \ref{TF_HCG_rc_ben_101}, it is clear that low mass galaxies are strongly influencing this result for the Tully-Fisher relation of compact groups. To confirm that, we removed galaxies having rotational velocities lower than 100 km s$^{-1}$. In this case, the Tully-Fisher relation for HCG galaxies is: $M_{\rm{B}}=(-4.26\pm1.73)-(7.07\pm0.74)\times[\rm{log}(\it{V}\rm{_{max}})]$ and for GHASP it is $M_{\rm{B}}=(-4.28\pm1.39)-(6.98\pm0.60)\times[\rm{log}(\it{V}\rm{_{max}})]$. We have also fitted an ordinary least square regression and a least  square bisector fit to the HCG and GHASP samples excluding objects  fainter than M$_{\rm{B}}$ = -18 and again we obtain very similar results, i.e. there are no significative differences in the slopes and zero points of the TF for the two samples. If we add GHASP and HCG galaxies (excluding HCG galaxies with rotational velocities lower than 100 km s$^{-1}$), we obtain: $M_{\rm{B}}=(-3.44\pm1.09)-(7.38\pm0.48)\times[\rm{log}(\it{V}\rm{_{max}})]$. This result is fully in agreement with the fit obtained for the GHASP sample. From this analysis we concluded that low mass galaxies in compact groups lie off the B-band Tully-Fisher relation while for galaxies with rotational velocities higher than 100 km s$^{-1}$, the TF relations for HCGs and GHASP galaxies have very similar slopes and zero points. 
If we take into account the error bars in the inclination and we remove
galaxies for which no error bars were quoted or having extremely large
error bars that produced errors in the computation (HCG 7c, 47d, 49a,
79d and 87a), we found that 75$\%$ of the HCG sample is inside 1$\sigma$ of the relation defined by GHASP (equation 1). A fraction of 25$\%$
of the compact group sample could not reach the Tully-Fisher relation
inside 1$\sigma$ (HCG 40e, 49b, 88c, 89d, 91c, 96c, 96d, 100c and
NGC92). Interestingly, HCG 49b, 89d, 96c, 96d and 100c have rotational
velocities lower than 100 km s$^{-1}$.

We study the dispersion that our samples present in the TF relation with respect to the fit derived from GHASP and HCG together (without the galaxies with v$<$100 km s$^{-1}$, in the case of HCG galaxies). We found that a total of 32$\%$ of the HCG and 25$\%$ of the GHASP galaxies are outside the 1$\sigma$ TF relation. This suggests a higher dispersion presented by the HCG galaxies although the main result here is that 98$\%$ and 97$\%$ of the HCG and GHASP samples are within the same relation at the 2$\sigma$ level, showing complete agreement at this level. The higher dispersion for HCGs could be due to the frequent interactions in this environment. Puech et al. (2008) found that galaxies at z$\sim$0.6 having a complex kinematics produced a high dispersion in the TF relation. This result is in agreement with the results of compact groups of galaxies.

In the low mass regime, a few field galaxies producing a high dispersion in the TF relation display lower B-band luminosities than the expected, and therefore, low stellar masses. This is due to the fact that the gaseous component is not negligible for low mass galaxies in the field.  If the gas mass is included, field low-mass galaxies are shifted towards the TF relation (as it was already shown in McGaugh et al. 2000). For HCG galaxies, however, adding the gas component moves their positions upwards, off the TF relation. The compact groups outliers have enhanced luminosities (probably due to star formation), which is not the case for field galaxies.  In this sense, it could be dangerous to quantify and compare the dispersions in the TF relation for HCG and GHASP galaxies at the low mass regime, given that the physical reason for such dispersion may be different.

\subsection{Interaction Indicators}

Mendes de Oliveira et al. (1998), Amram et al. (2003) and Plana et
al. (2003) have devised a scheme to classify galaxies as interacting or
not, based on the analysis of 2-D velocity fields and other indicators. Using this same scheme, Torres--Flores et
al. (2009) classified HCG 2, HCG 7, HCG 22 and NGC 92 in different
evolutionary stages. In a similar way, we classified the galaxies studied here as
interacting or not. In Table \ref{table5} we show the main interaction
indicators. Of all 11 galaxies studied in this new sample (from which we have derived rotation curves for 9 of them) 6 present at least two interaction indicators and can be classified as interacting. 

\begin{table}
\begin{minipage}[h!]{\columnwidth}
\caption{Interaction Indicators}
\label{table5}
\centering
\renewcommand{\footnoterule}{}  
\begin{tabular}{ccccccc}
\hline \hline
\multicolumn{1}{c}{Interaction\footnotetext{Interaction indicators for the new sample presented in this paper. (1): Highly disturbed velocity field. (2): Disagreement between both sides of the rotation curve. (3): Gaseous versus stellar major-axis misalignment. (4): Changing position angle along major axis. (5): Tidal Tails. (6): High IR luminosity. (7): Central activity. The symbol ``+'' means that the indicator is ``on'' otherwise the symbol is ``--''.}} &  \multicolumn{6}{c}{Galaxy (HCG)} \\ 
indicator & 37d & 40c & 40e & 47a & 47d &49a \\
\hline
(1) & -- & -- & -- & -- & --  & --   \\
(2) & +  & -- & +  & ...& ... & +   \\
(3) & -- & -- & +  &  + &  +  & --  \\
(4) & -- & -- & -- & -- & --  & +   \\
(5) & -- & -- &    & -- & --  & +     \\
(6) & -- & -- & +  & +  & --  & ... \\
(7) & -- & -- & -- & -- & ... & ... \\
\hline
         & 49b & 49c & 56a & 68c & 93b  \\
\hline
(1) & + & -- & -- & -- & +  &   \\
(2) & + & +  & -- & -- & +  &  \\
(3) & + & +  & -- & -- & +  &   \\
(4) & + & -- & -- & -- & -- &   \\
(5) & --& -- & -- & -- & -- &   \\
(6) &...& ...& -- & ...& ...&  \\
(7) &...& ...& -- & +  & -- & \\
\hline
\end{tabular}
\end{minipage}
\end{table}

One of the interaction parameters listed in Table \ref{table5} is the difference
between the kinematically and morphological-derived position angles.
We plot this difference for 46 compact group galaxies 
in Fig. \ref{pa} (bottom panel). We note that this 
difference is an indicator of interactions only for
non-barred galaxies given that it is common to find bars which are not
aligned with kinematic major axes of galaxies. If we discard the barred galaxies we
still find that almost one third of the sample of 32 galaxies studied in detail
so far (this paper, Mendes de Oliveira et al. 2003 and Torres-Flores et
al. 2009) has a misalignment larger than 20 degrees between the kinematic and the morphological
position angle, yielding yet more pieces of evidence for the high incidence of interactions
among galaxies in compact groups. We noted that only one galaxy (HCG 49a) has a kinematic and morphological inclination lower than 25 degrees. This object has a small misalignment between kinematic and morphological PA (7 degrees), therefore, its low inclination is not affecting the determination of the position angles of this galaxy. 

A new interaction indicator for the whole sample of compact groups could be associated with the number of galaxies for which it is possible to obtain the rotation curves versus the total number of observed galaxies. If we gather all the HCG galaxies that have Fabry-Perot observations, discarding elliptical and S0, we found that rotation curves could be obtained for 68$\%$ of the observed galaxies. In the case of the GHASP sample, for 85$\%$ of the observed galaxies it is possible to derive their rotation curves. This fact could be associated with the complex kinematics shown by HCG galaxies. Moreover, HCG presents a deficiency in the content of HI gas (Verdes-Montenegro et al. 2001), therefore we can not exclude the lack of gas in the HCG galaxies as an important bias in this comparison. 

\section{Summary and Conclusions}

We have presented new velocity fields, monochromatic H$\alpha$ images and velocity dispersion maps for galaxies in nine compact groups. We have explored the B-band Tully-Fisher relation for a sample of 41 galaxies in compact groups and we compared it with the reanalyzed Tully-Fisher relation for the GHASP sample. We found that galaxies in dense environments lie on the Tully-Fisher relation defined by field galaxies, however, some of the low-mass galaxies have either a high luminosity for their mass or a high rotational velocity for their luminosities (e.g. HCG 49b). We also found that one third of the non-barred galaxies in compact groups have a misalignment between the kinematic and morphological position angle higher than 20 degrees.

\begin{acknowledgements}
We would like to thank the anonymous referee for very useful comments and suggestions which were very important in improving this paper. S. T-F acknowledges the financial support of FAPESP through a
Ph.D. fellowship (2007/07973-3). S. T-F also acknowledges the financial support of Egide through a Eiffel scholarship. CMdO acknowledgs support from FAPESP
(2006/56213-9) and CNPq. HP acknowledges financial support from CAPES (3656/08-0).  S. T-F, CMdO and HP would like to thank the staff members of the Laboratoire d'Astrophysique de Marseille for their hospitality when part of this work was developed. This research has made use of the NASA/IPAC Extragalactic Database (NED) which is operated by the Jet Propulsion Laboratory, California Institute of Technology, under contract with the
National Aeronautics and Space Administration. We also acknowledge the
use of the HyperLeda database (http://leda.univ-lyon1.fr).
\end{acknowledgements}

\Online

\begin{appendix} 

\section{Individual galaxy properties} 

\subsection{HCG 37}

This group is formed by five galaxies. The brighter member, HCG 37a, has a morphological type E7 (Hickson 1992). Coziol et al. (2004) classified galaxies HCG 37a, HCG 37b, HCG 37c and HCG 37d as a dwarf LINER (dLINER), LINER, low luminosity AGN and a Star Forming Galaxy (SFG), respectively. HCG 37b is a near IR source (Allam et al. 1996). Verdes-Montenegro et al. (2001) reported that 87\% of the expected HI is missing in HCG 37. Four of the five galaxies of HCG 37 were observed (due to the field of view of the observations) and we detect H$\alpha$ emission only in two of them.
HCG 37a shows a central H$\alpha$ emission, as was reported in Vilchez et al. (1998), however, the H$\alpha$ profiles have a low signal to noise ratios. For this galaxy it was not possible to perform a rotation curve.

The monochromatic map of HCG 37d shows an extension of 13 arcsec (5.2 kpc). It has two peaks in the emission, as was found in Vilchez et al. (1998), the peak located in the north west side of the galaxy being stronger. The continuum image shows the same peaks as the monochromatic image. The velocity field of this object is regular. The kinematic position angle is aligned with the optical one (as derived by Mendes de Oliveira 1992 at 2 \textit{h$^{-1}$}kpc), differing in only 2$^{\circ}$. The rotation curve of HCG 37d is not symmetric and both sides do not match. The receding and approaching sides reach velocities of 60 km s$^{-1}$ and 53 km s$^{-1}$ respectively. The rotation curve does not reach the optical radius of the galaxy (R$_{25}$).

\subsection{HCG 40}

This quintet is composed by two early-type and three late-type galaxies. HCG 40a, HCG 40d and HCG 40e were classified as radio sources (dLNR, LNR and Sy2 respectively) in Coziol et al. (1998) and re-classified as no-emission galaxy, SFG and SFG in Coziol et al. (2004). HCG 40b is a no-emission galaxy and HCG 40c is a SFG (Coziol et al. 2004). HCG 40e is a FIR source (Allam et al. 1996). In the evolutive scenario  of Verdes-Montenegro et al. (2001), HCG 40 is a type 3a group, where 89\%  of the expected HI mass is missing, which convert this group in a very evolved group. 

The strongest H$\alpha$ emission of the group is detected in two peaks, in the barred early-type spiral HCG 40d. This emission is very concentrated in the central region, probably associated with star formation at the end of the bar. We could not derived the rotation curve for this object.

We also detected H$\alpha$ emission in the central region of HCG 40a and HCG 40b (both galaxies were classified as early-type galaxies by Hickson (1992), E3 and S0, respectively). Interestingly, no emission was detected in Coziol et al. (2000) for these objects. As for HCG 40d, we could not obtain the rotation curve for these objects.

HCG 40c displays an extended monochromatic map. We detected a peak in the H$\alpha$ emission at the south-east side of HCG 40c, coincident with the peak in the HI distribution shown in Verdes--Montenegro et al. (2001). Its velocity field is regular. The kinematic position angle does not change with radius. The rotation curve is symmetric, reaching a plateau at 6 kpc, however, this curve does not reach the optical radius R$_{25}$. At radii larger than 10 kpc, there is no more H$\alpha$ emission, as it is possible to note from the monochromatic image.

HCG 40e shows a weak H$\alpha$ emission on its disk. Both sides of the rotation curve do not match.

\subsection{HCG 47}

This quartet is formed by two pairs of galaxies. HCG 47a is a SBb (Hickson 1993) galaxy, showing a ring-shape structure on its south west side (Fasano et al. 1994). Shimada et al. (2000) classified HCG 47a as an absorption-line galaxy. Hickson classified HCG 47b, HCG 47c and HCG 47d as E3, Sc and Sd galaxies. HCG 47a and HCG 47b are radio and FIR sources (Allam et al. 1996). 

We detected a strong H$\alpha$ emission in the ring of HCG 47a. On the other hand, HCG 47b shows weak intensity in its monochromatic map. The north side of HCG 47d shows the strongest H$\alpha$ emission in this galaxy. However, we also note an extended emission in its disk. HCG 47c only shows a central knot in its monochromatic image, at the same position of the continuum peak.

The velocity field shows higher velocities in the eastern part of the approaching side than in the western region, probably due to the presence of the ring. The kinematic position angle differs from the optical one, at 2\textit{h$^{-1}$}kpc and at R$_{25}$, in 14$^{\circ}$ in both cases. 

HCG 47d shows a regular velocity field. However, there is an offset between the kinematic and optical position angle of 32$^{\circ}$ and 23$^{\circ}$ (Mendes de Oliveira 1992 and HyperLeda, respectively). For HCG 47a and 47d it was not possible to obtain the rotation curves (the procedure developed by Epinat et al. 2008a does not converge for these galaxies).

\subsection{HCG 49}

This small and compact quartet of galaxies shows an angular diameter of 0.9 arcmin (Hickson et al. 1982) (36 kpc) and is the most distant group in this sample (140 Mpc). HCG 49a and HCG 49b are late-type spiral galaxies, HCG 49c is a Im and HCG 49d is a E5 (Hickson 1993). In the scenario proposed in Verdes--Montenegro et al. (2001), HCG 49 is a well developed group, in the 3b phase, where the HI gas forms a large cloud containing all galaxies, with a single velocity gradient.

Fasano et al. (1996) noted that HCG 49a seems to be a superposition of two galaxies but our monochromatic map shows that the emission is coming from only one object. We analyzed the H$\alpha$ profile in the region where  the superposition appears, and we do not detect a defined profile. The monochromatic map of HCG 49a shows H$\alpha$ emission across the disk of this object. This emission peaks at the same place in the monochromatic and the continuum map. This object shows a regular velocity field. Kinematic position angle differs only in 7 $^{\circ}$ from the optical position angle at 2 \textit{h$^{-1}$}kpc. However, the difference is larger if we take the optical position angle at R$_{25}$ (16 $^{\circ}$). The rotation curve is symmetric and both sides match relatively well. 

HCG 49b shows two strong peaks in the monochromatic map. One of them is located at the same position of the continuum maximum. The other bright knot is placed to the east of the galaxy. This knot shows an amplitude of $\sim$40 km s$^{-1}$. HCG 49b shows a perturbed velocity field. The kinematic position angle changes along the radius and it differs from the optical one at 2 \textit{h$^{-1}$}kpc in 17$^{\circ}$. However, at R$_{25}$ the optical and kinematic position angles reach a similar value (Table \ref{table3}). The rotation curve of HCG 49b extended out to 12.5 arcsec($\sim$8.5 kpc). Both sides do not match. The receding side reaches values of 70 km s$^{-1}$ and the approaching side reaches values of 75 km s$^{-1}$.

The monochromatic map of HCG 49c shows a much lower intensity than HCG 49a and b. We detected a bridge in the H$\alpha$ emission between members b and c. The velocity field seems regular. However, a disagreement of 64$^{\circ}$ is observed between the kinematic and optical position angles. Both sides of the rotation curve match from the center out to 2.5 kpc. Beyond this radius, no H$\alpha$ emission is detected in the approaching side.

\subsection{HCG 54}

Hickson et al. (1989) classified this group as formed by three irregular galaxies and one Sdm galaxy. Verdes-Montenegro et al. (2002) found HI tidal tails and shells in the environment of HCG 54. They concluded that this group is a merger remnant.

From our monochromatic map we note that HCG 54b shows the strongest H$\alpha$ emission and its velocity field shows an amplitude of 30 km s$^{-1}$. HCG 54b and 54d also appear in the monochromatic map, however it is difficult to get some information from the velocity field. HCG 54a is detected in the continuum image. Our data supports the scenario that HCG 54 is in fact one only galaxy and not a group. Due to the nature of this object (Verdes-Montenegro et al. 2002) we decided do not obtain any kinematic parameter of its members.

\subsection{HCG 56}

This quintet if formed by the late-type spiral HCG 56a, the SB0 HCG 56b and three SO galaxies (c, d, e). Coziol et al. (2004) classified members a, d and e as Star Forming Galaxies, HCG 56b as a Seyfert galaxy and HCG 56c as a no-emission line galaxy. Allam et al.(1996) detected members b,c and d at 25 60 and 100 $\mu$m. This group only shows 19\% of its expected HI mass (Verdes--Montenegro et al. 2001). We present Fabry-Perot data for members a,b,c and d of this group.
The monochromatic map of HCG 56a shows an extended emission along the disk. 
Peaks in the H$\alpha$ emission are detected in the southern and northern side of this galaxy. Both peaks are located at the same distance with respect to the peak in the continuum image. The velocity field of HCG 56a seems to be regular. We noted that the approaching side is more extended than the receding side. We do not detect any misalignment  between the kinematic and optical position angles for this galaxy. The rotation curve of HCG 56a is symmetric. The approaching and receding sides match from  3 to 18$\arcsec$ (2 to 10 kpc), where the approaching side reaches a plateau. At radii larger than 18$\arcsec$, there is no data for the receding side, however, the approaching side reachs 35$\arcsec$.

We detected H$\alpha$ emission at the center of the S0 galaxies HCG 56b, HCG 56c and HCG 56d. Given that we only detected H$\alpha$ emission at the center of these objects, it was not possible to obtain their rotation curves. No emission was detected in the bridge between members b and c. Members b, c and d shows a shift of 0.88$\arcsec$ (one pixel) between their monochromatic and continuum peaks.

\subsection{HCG 68}

This group is formed by four early-type galaxies (HCG 68a, b, d and e) and one barred spiral galaxy (HCG 68c) (Hikcson 1993). Shimada et al. (2000) classified members a, b and c as AGN, however, Coziol et al. (2004) classified HCG 68a as a no-emission galaxy. Verdes-Montenegro et al. (2001) detected only 33\% of the expected HI mass for this group.
We detected H$\alpha$ emission in the monochromatic map of HCG 68a and b. However, the H$\alpha$ profile is not well defined in both galaxies. 

We detected several bright HII regions on the well defined arms of HCG 68c. On the nuclear region, we detected a peak in H$\alpha$, nevertheless, no emission was detected across the bar of this object, as also reported by Vilchez et al. (1998). The maximum intensity in the monochromatic map of HCG 68c corresponds to an HII region located 60$\arcsec$ from its center (north west side of the galaxy). The velocity field of HCG 68c is quite regular. The kinematic and optical (R$_{25}$) position angles are aligned. The difference between the kinematic and optical position angle at 2\textit{h$^{-1}$} is caused by the influence of the bar at this radius. The rotation curve of HCG 68c reaches 115$\arcsec$ (19 kpc). In the first 10$\arcsec$, both sides of the rotation curve do not match and we detected a bump in the approaching side, reaching 250 km s$^{-1}$. Both sides of the rotation curve match from 15 to 105$\arcsec$. At 40$\arcsec$, the approaching side shows higher velocities than the receding side, showing some bumps, probably produced by the crossing of spiral arms. 

\subsection{HCG 79}

Mendes de Oliveira et al. (2003) presented the rotation curve of HCG 79d. Inclinations for HCG 79d are 64$^{\circ}$ and 78$^{\circ}$ (depending if it is taken from the axis ratio given in Hickson 1993 or if it is taken from HyperLeda database). Durbala et al. (2008) showed continuum-subtracted H$\alpha$ images of HCG 79 (for this group, see details in Durbala et al. 2008).

\subsection{HCG 93}

This quartet is formed by two barred spiral galaxies, HCG 93b and HCG 93c, one E1 galaxy (HCG 93a) and the SB0 galaxy HCG 93d (Hickson 1993). Members a and c were classified as AGN by Shimada et al. (2000) and Coziol et al. (2004). HCG 93b is classified as an AGN by Shimada et al. (2004) but Coziol et al. (2004) classified it as a Star Forming Galaxy. About 84\% of the expected HI mass in HCG 93 is missing (Verdes-Montenegro et al. 2001). 
We presented Fabry-Perot data only for the barred spiral HCG 93b. Its monochromatic map shows H$\alpha$ emission across the bar and two bright knots at the end of it, produced by star formation resulting from the compression of gas in this place, in agreement with its morphological type (Phillips 1996, Pisano et al. 2000). In the arms, we detected a few HII regions. The velocity field of HCG 93b shows a twist. We detected a disagreement between optical and kinematic position angles of 17$^{\circ}$ and 134$^{\circ}$ (Mendes de Oliveira et al. 2003 and HyperLeda, respectively). Both sides of the rotation curve of HCG 93b do not match. From 20$\arcsec$ (7 kpc), the approaching side falls, reaching values of $\sim$120 km s$^{-1}$. The approaching side has an extension of 95$\arcsec$ (30 kpc).

\end{appendix}

\begin{appendix} 

\section{Individual maps and rotation curves}

\begin{figure}[h]
\includegraphics[width=\textwidth]
{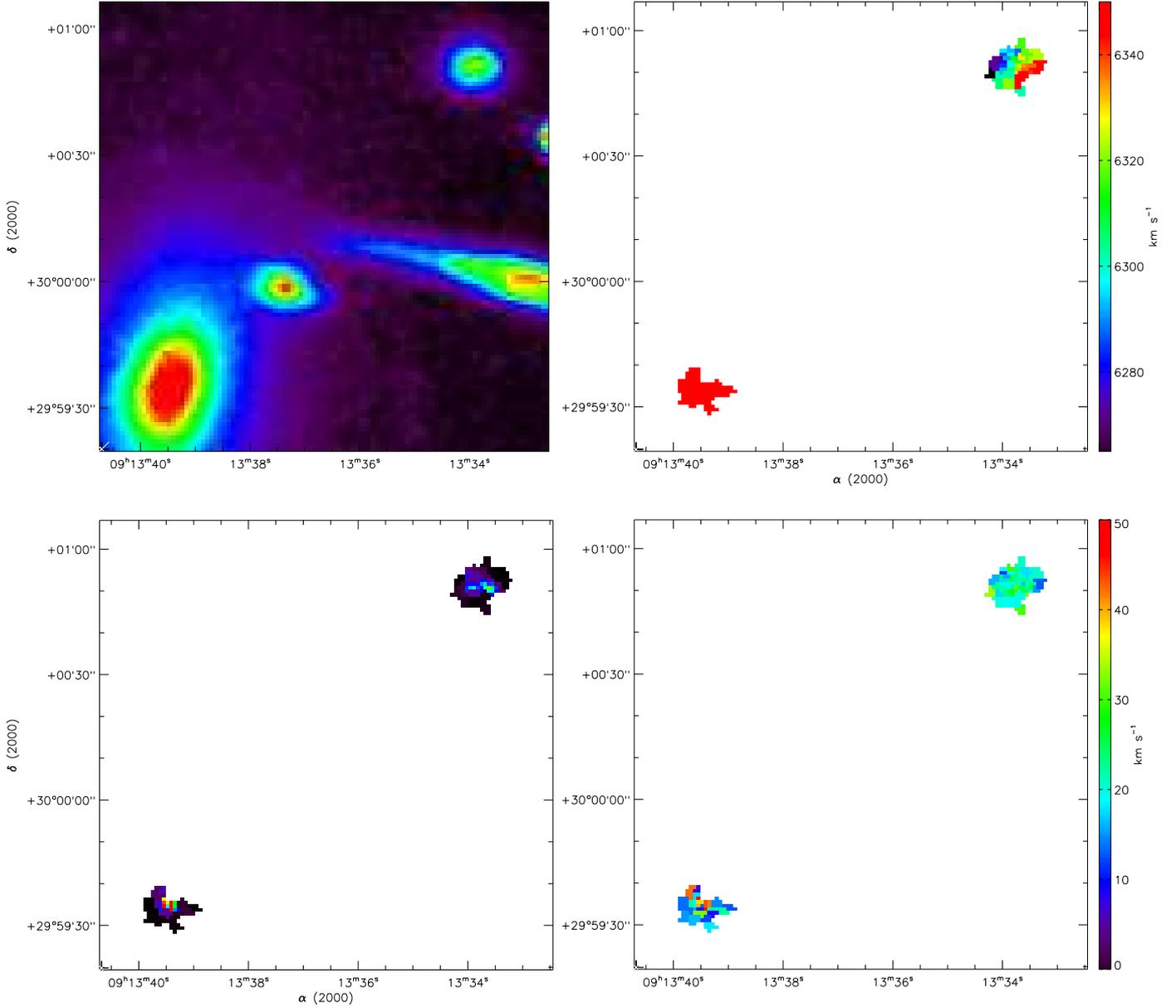}
\caption{Maps for HCG 37. North is at the top and east to the left for all images in Figs. \ref{maps_hcg37d} to \ref{maps_hcg93b}. HCG 37a is located to the south east of HCG 37d. Top left: B band image from DSS. Top right: Velocity field. Bottom left: Monochromatic image. Bottom right: velocity dispersion map.}
\label{maps_hcg37d}
\end{figure}
\clearpage

\begin{figure*}[t!]
\includegraphics[width=\textwidth]
{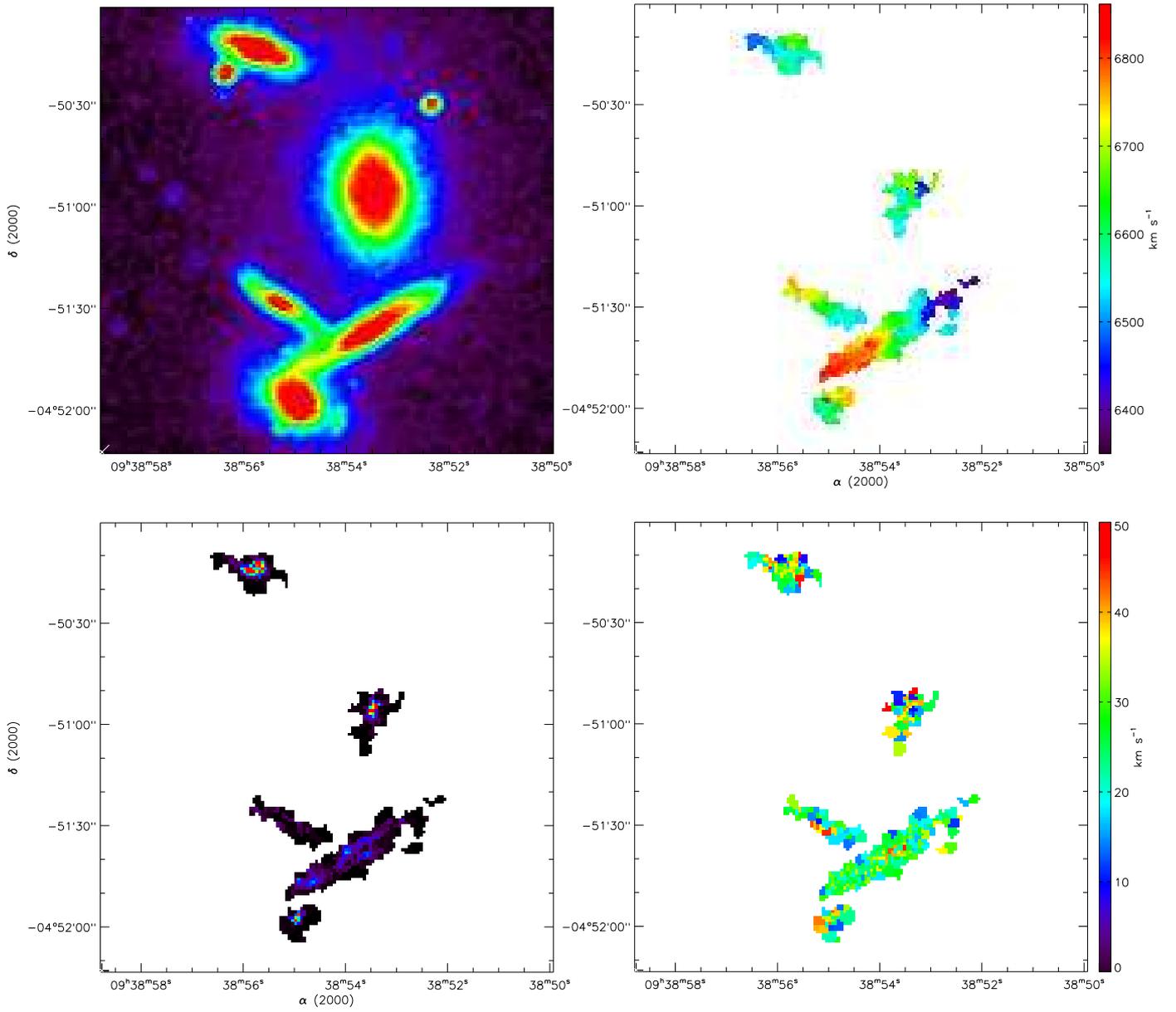}
\caption{Maps for HCG 40. HCG 40a is located close to the center of the field and to the south west of HCG 40d. HCG 40e, b and c are located at the south-east of HCG 40a. Top left: B band image from DSS. Top right: Velocity field. Bottom left: Monochromatic image. Bottom right: velocity dispersion map.}
\label{maps_hcg40}
\end{figure*}
\clearpage

\begin{figure*}[t!]
\includegraphics[width=\textwidth]
{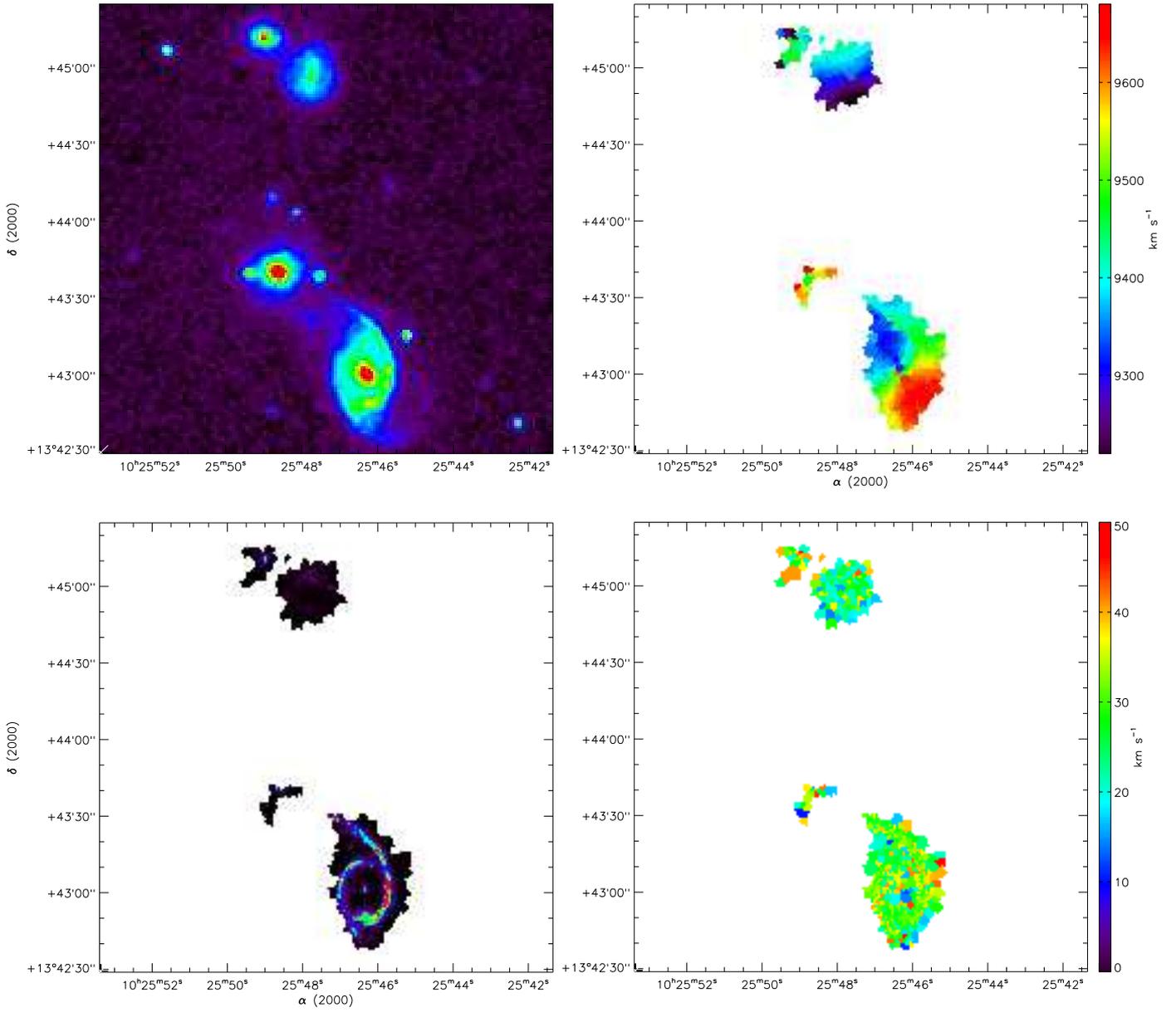}
\caption{Maps for HCG 47. HCG 47b and 47a are located at the bottom of the field. HCG 47c and 47d are located at the north of the field. Top left: B band image from DSS. Top right: Velocity field. Bottom left: Monochromatic image. Bottom right: velocity dispersion map.}
\label{maps_hcg47}
\end{figure*}
\clearpage

\begin{figure*}[t!]
\includegraphics[width=\textwidth]{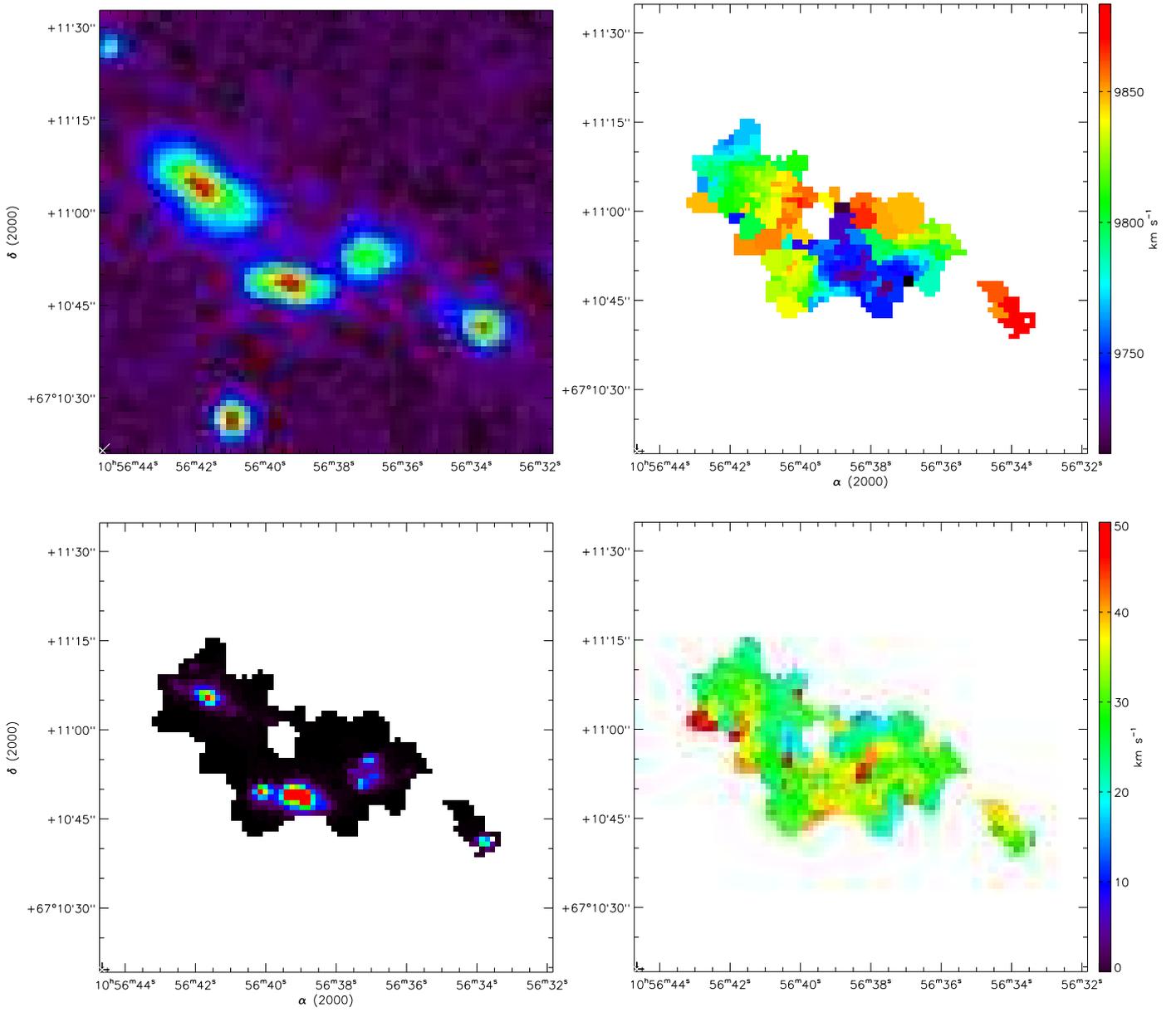}
\caption{Maps for HCG 49. From east to west: HCG 49a, 49b, 49c and 49d. Top left: B band image from DSS. Top right: Velocity field. Bottom left: Monochromatic image. Bottom right: velocity dispersion map.}
\label{maps_hcg49}
\end{figure*}
\clearpage

\begin{figure*}[t!]
\includegraphics[width=\textwidth]{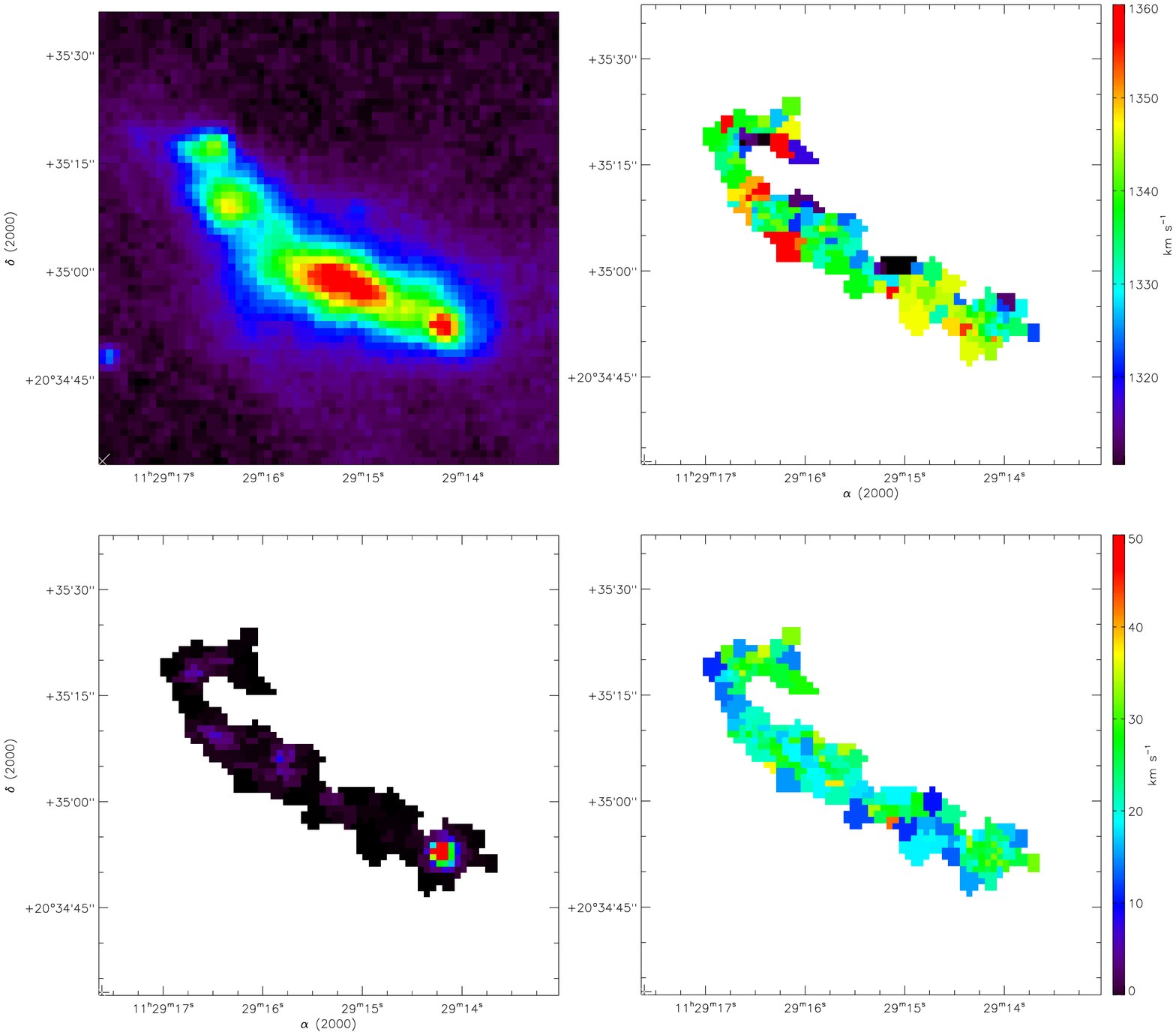}
\caption{Maps for HCG 54. Top left: B band image from DSS. Top right: Velocity field. Bottom left: Monochromatic image. Bottom right: velocity dispersion map.}
\label{maps_hcg54}
\end{figure*}
\clearpage

\begin{figure*}[t!]
\includegraphics[width=\textwidth]{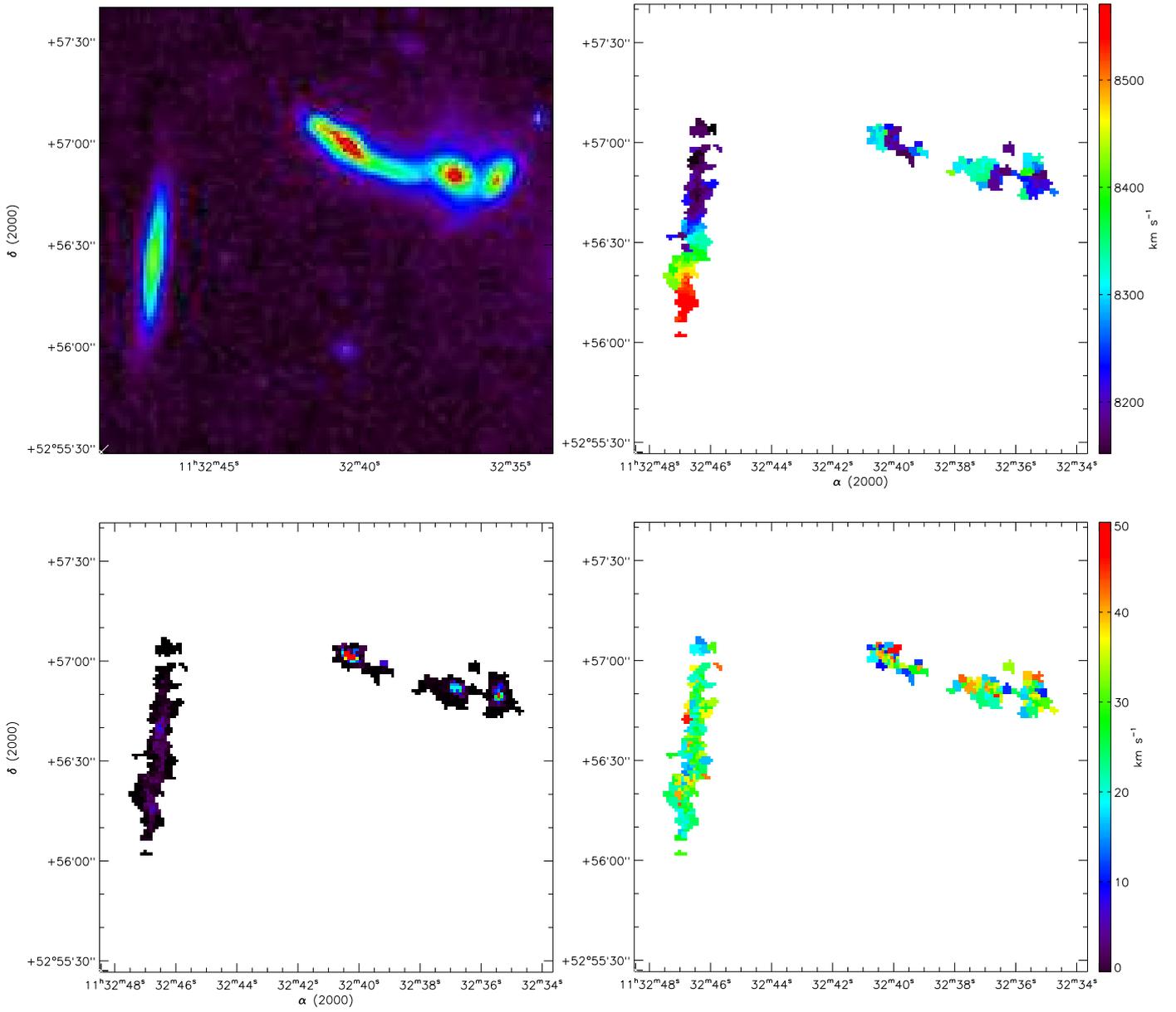}
\caption{Maps for HCG 56. From east to west: HCG 56a, 56b, 56c and 56d. Top left: B band image from DSS. Top right: Velocity field. Bottom left: Monochromatic image. Bottom right: velocity dispersion map.}
\label{maps_hcg56}
\end{figure*}
\clearpage

\begin{figure*}[t!]
\includegraphics[width=\textwidth]{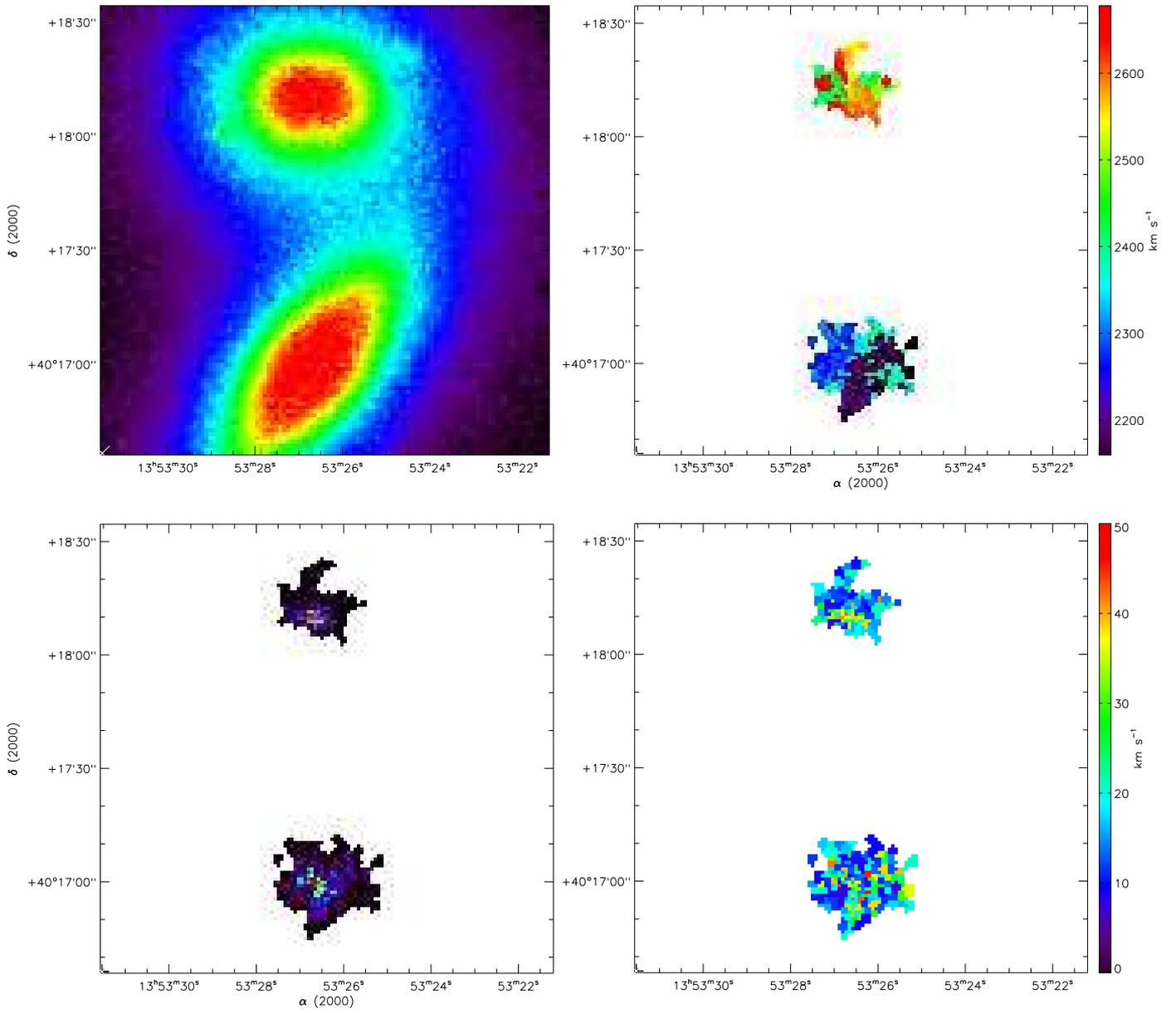}
\caption{Maps for HCG 68. HCG 68a is located to the south of HCG 68b. Top left: B band image from DSS. Top right: Velocity field. Bottom left: Monochromatic image. Bottom right: velocity dispersion map.}
\label{maps_hcg68ab}
\end{figure*}
\clearpage

\begin{figure*}[t!]
\includegraphics[width=\textwidth]{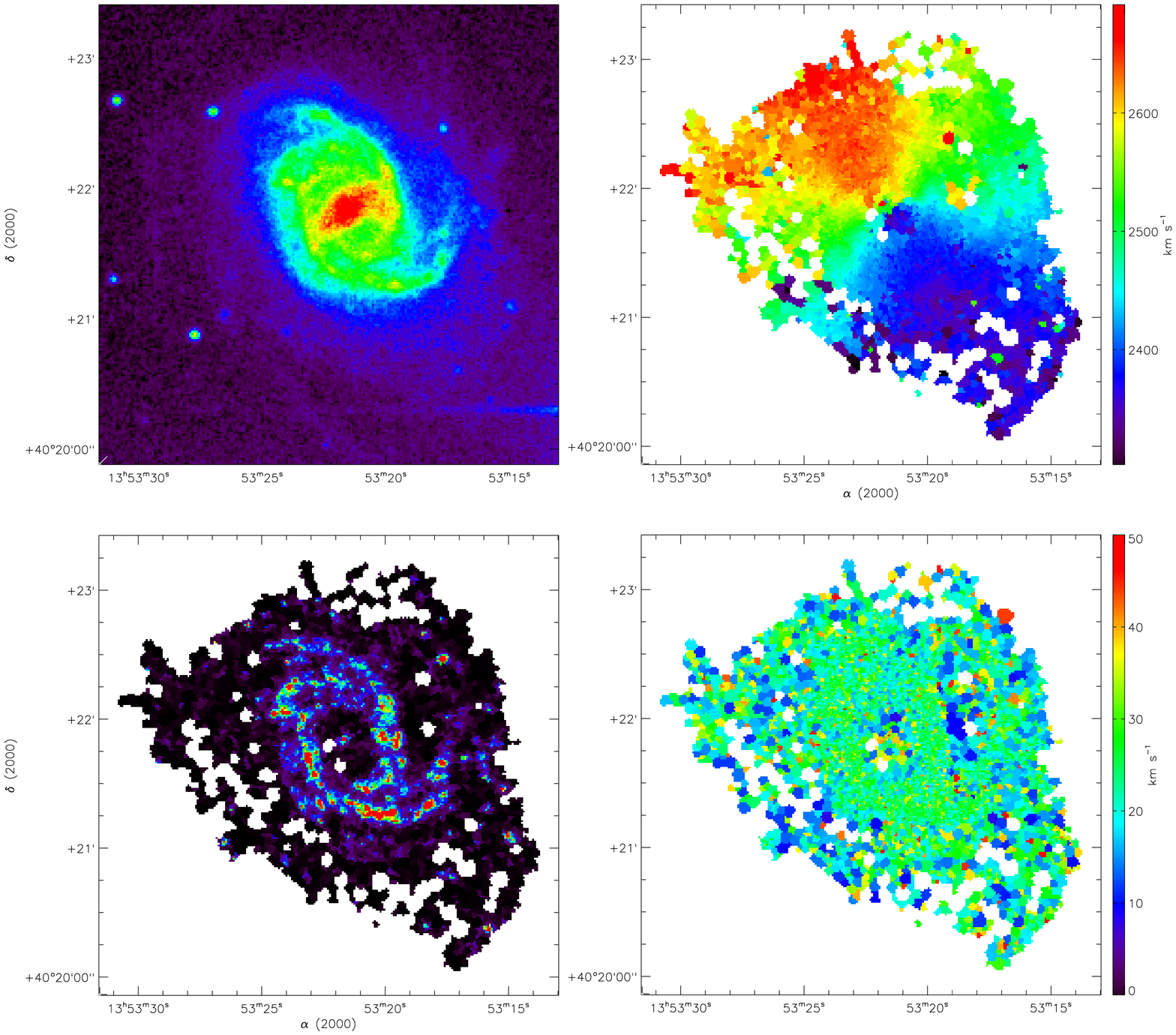}
\caption{Maps for HCG 68c. Top left: B band image from DSS. Top right: Velocity field. Bottom left: Monochromatic image. Bottom right: velocity dispersion map.}
\label{maps_hcg68c}
\end{figure*}
\clearpage

\begin{figure*}[t!]
\includegraphics[width=\textwidth]{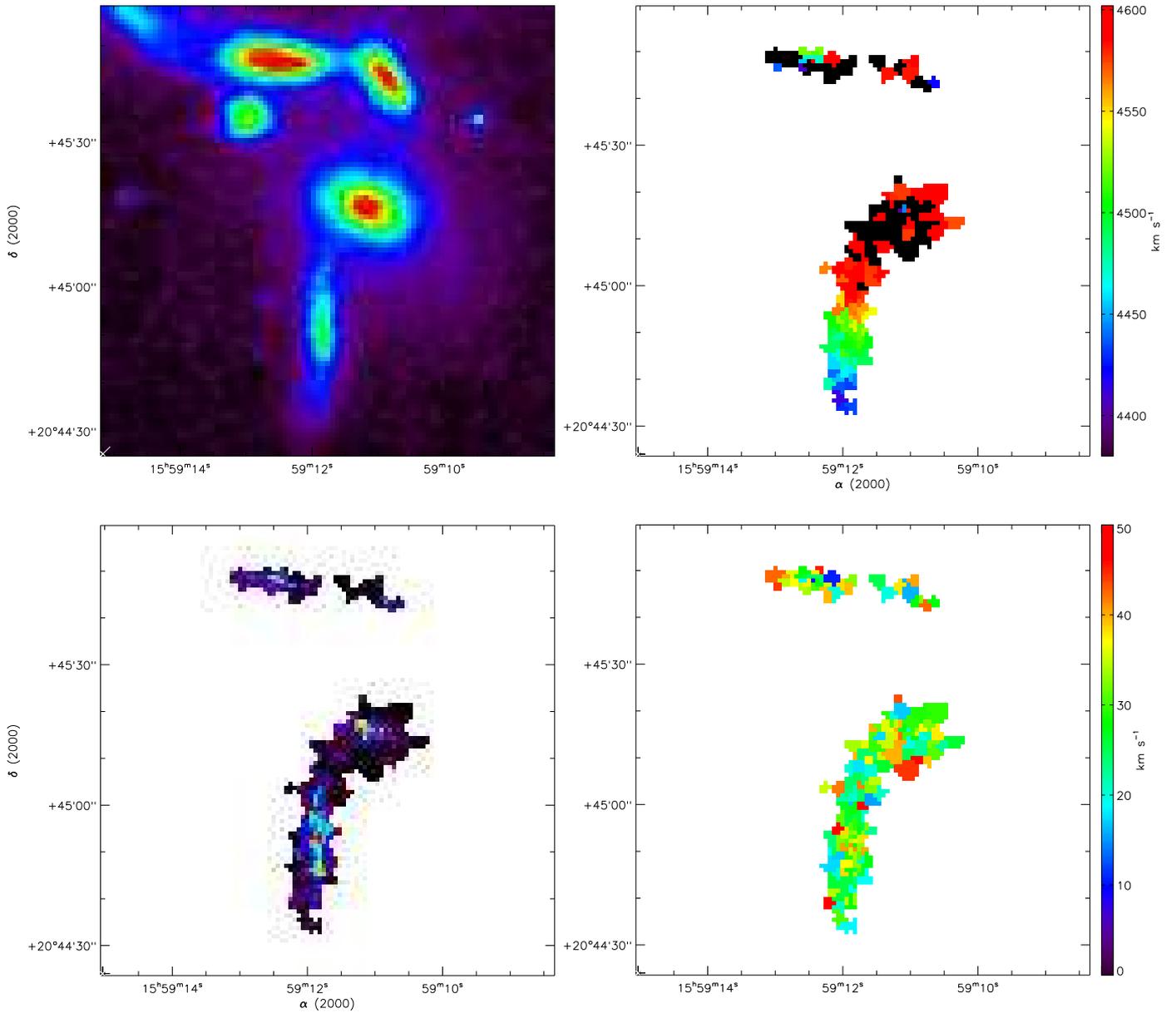}
\caption{Maps for HCG 79. HCG 79b and 79c are located in the north of the field, from east to west. HCG 79a is in the center of the field and HCG 79d in placed at the south of 79a. Top left: B band image from DSS. Top right: Velocity field. Bottom left: Monochromatic image. Bottom right: velocity dispersion map.}
\label{maps_hcg79}
\end{figure*}
\clearpage

\begin{figure*}[t!]
\includegraphics[width=\textwidth]{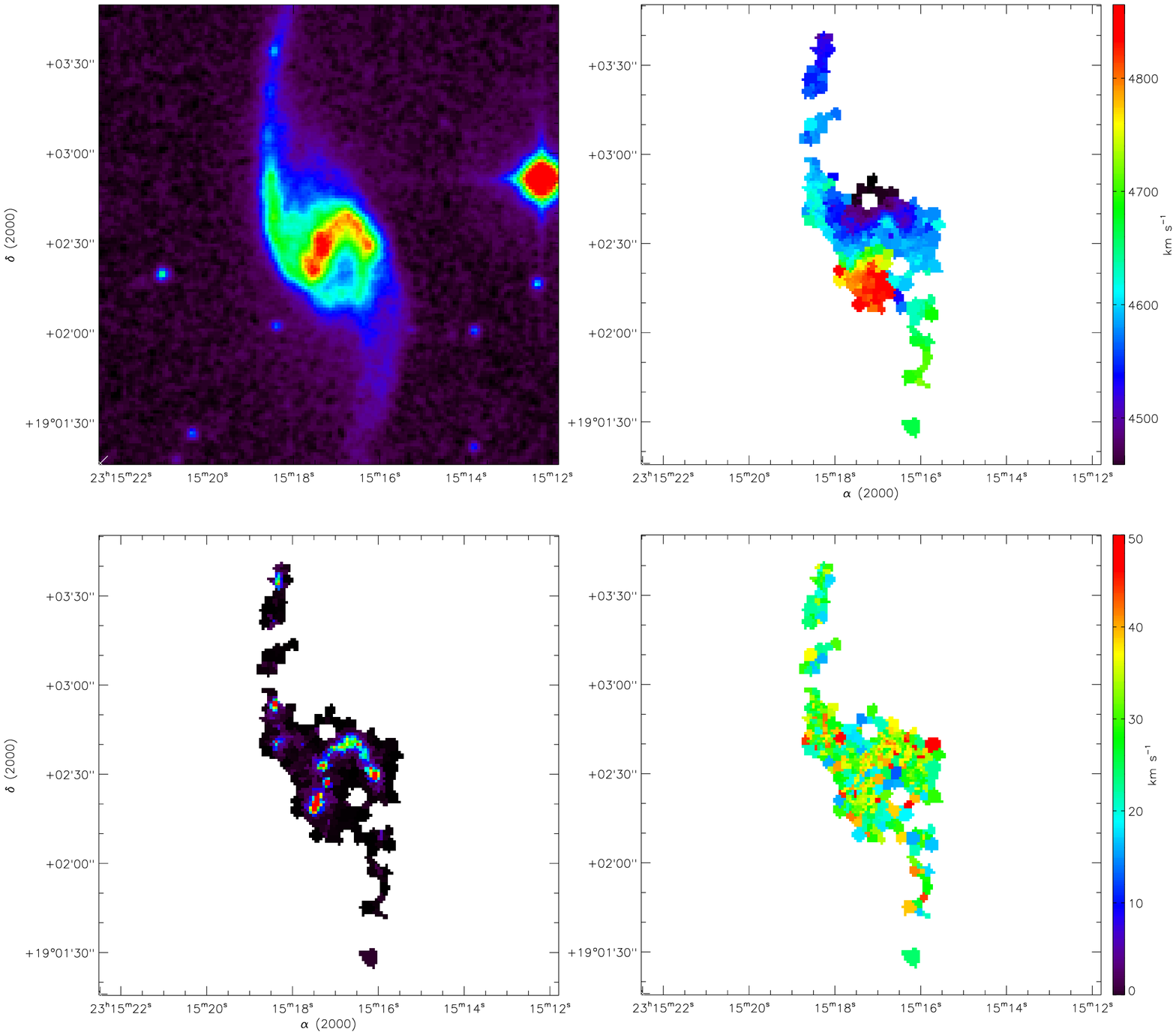}
\caption{Maps for HCG 93b. Top left: B band image from DSS. Top right: Velocity field. Bottom left: Monochromatic image. Bottom right: velocity dispersion map.}
\label{maps_hcg93b}
\end{figure*}
\clearpage

\begin{figure*}[h!]
\includegraphics[width=\columnwidth]{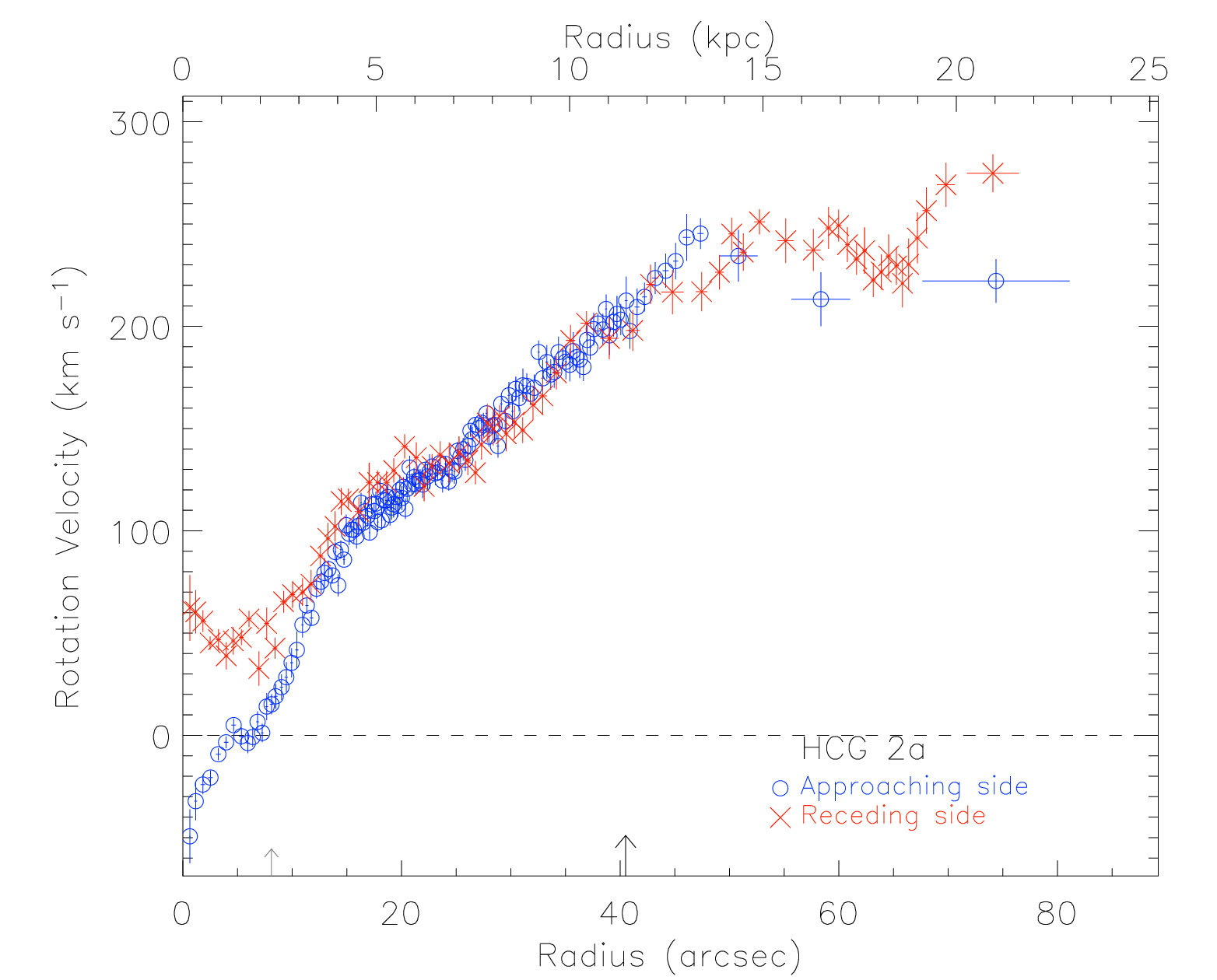}\includegraphics[width=\columnwidth]{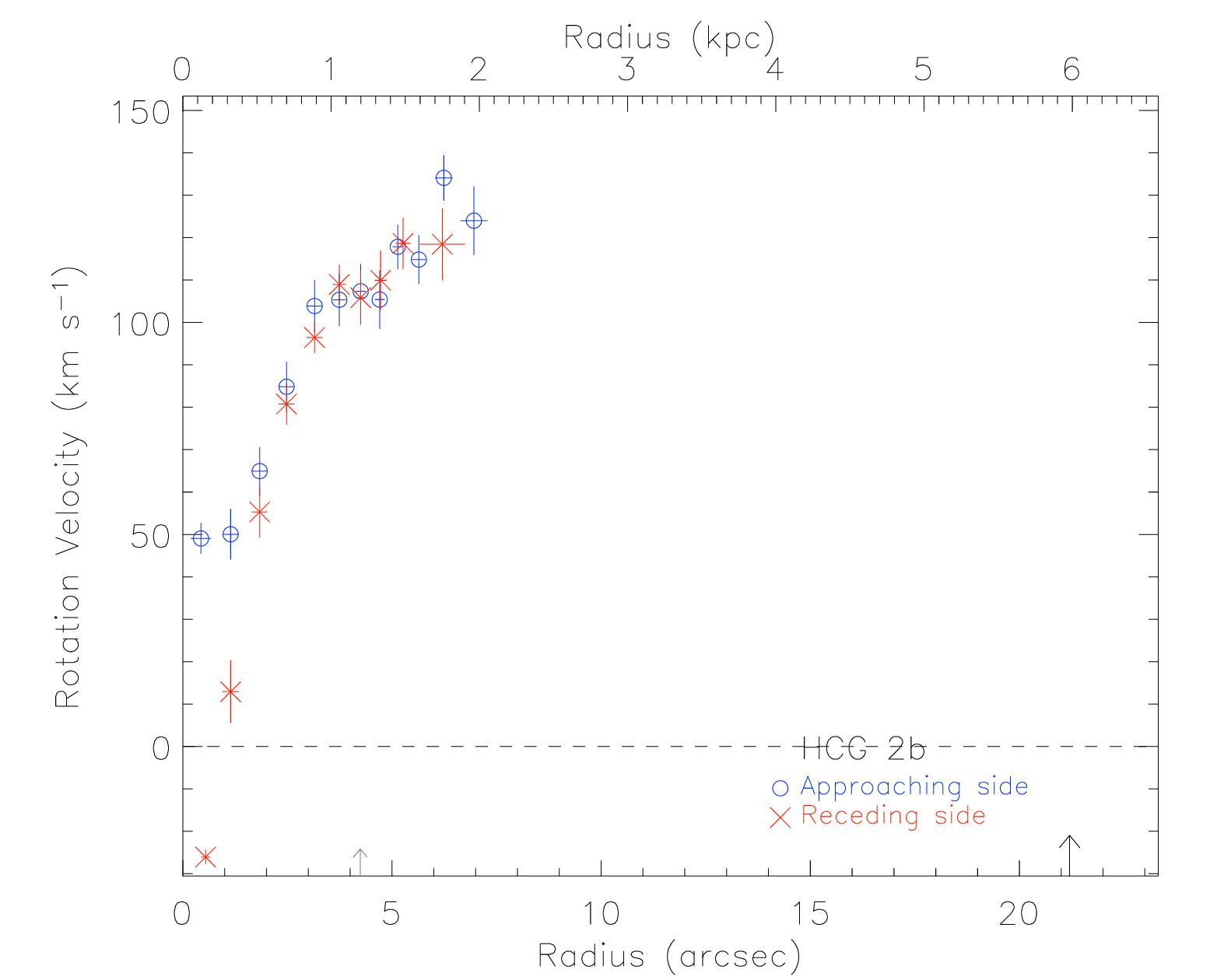} 
\end{figure*}

\begin{figure*}[h!]
\includegraphics[width=\columnwidth]{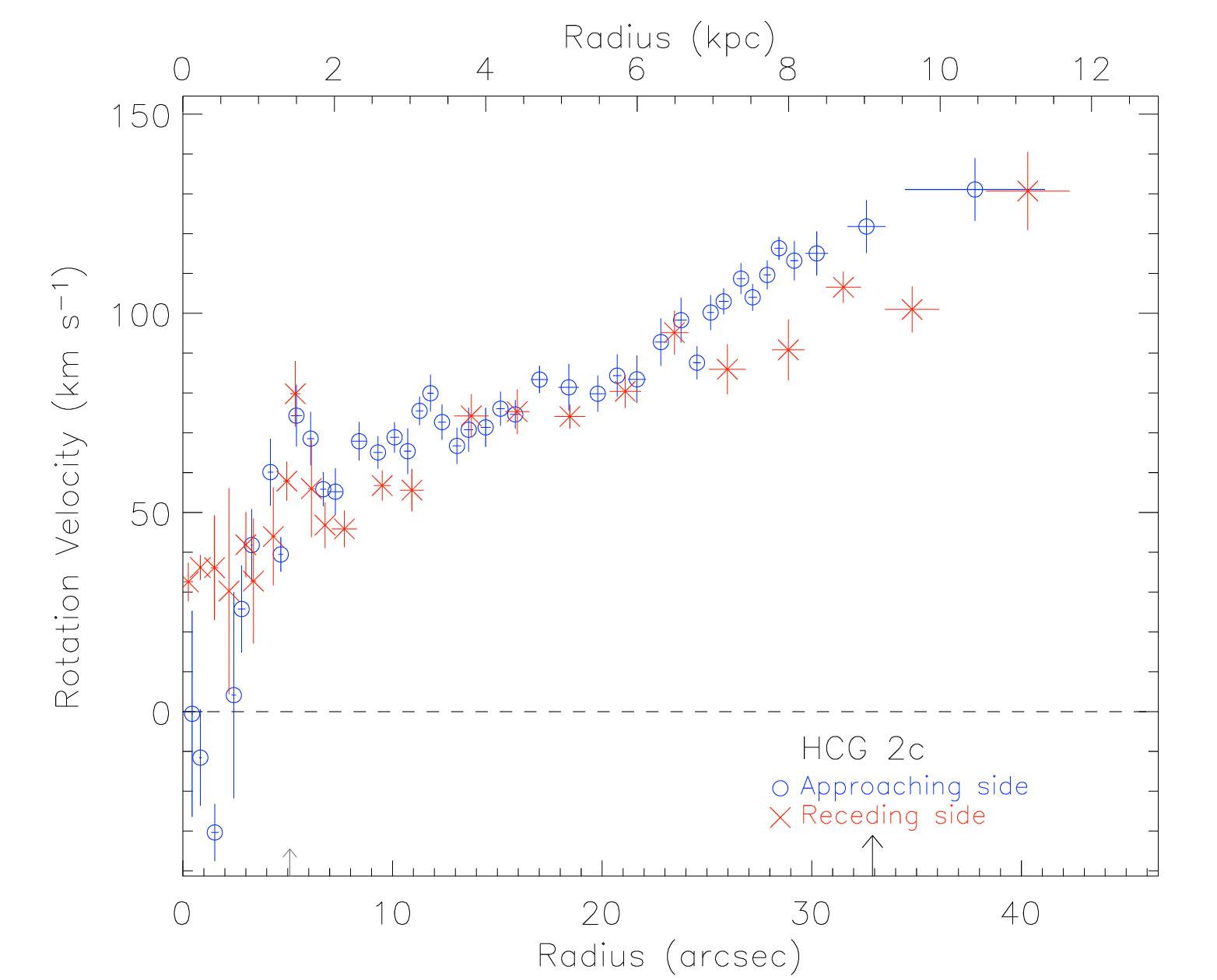}
\includegraphics[width=\columnwidth]{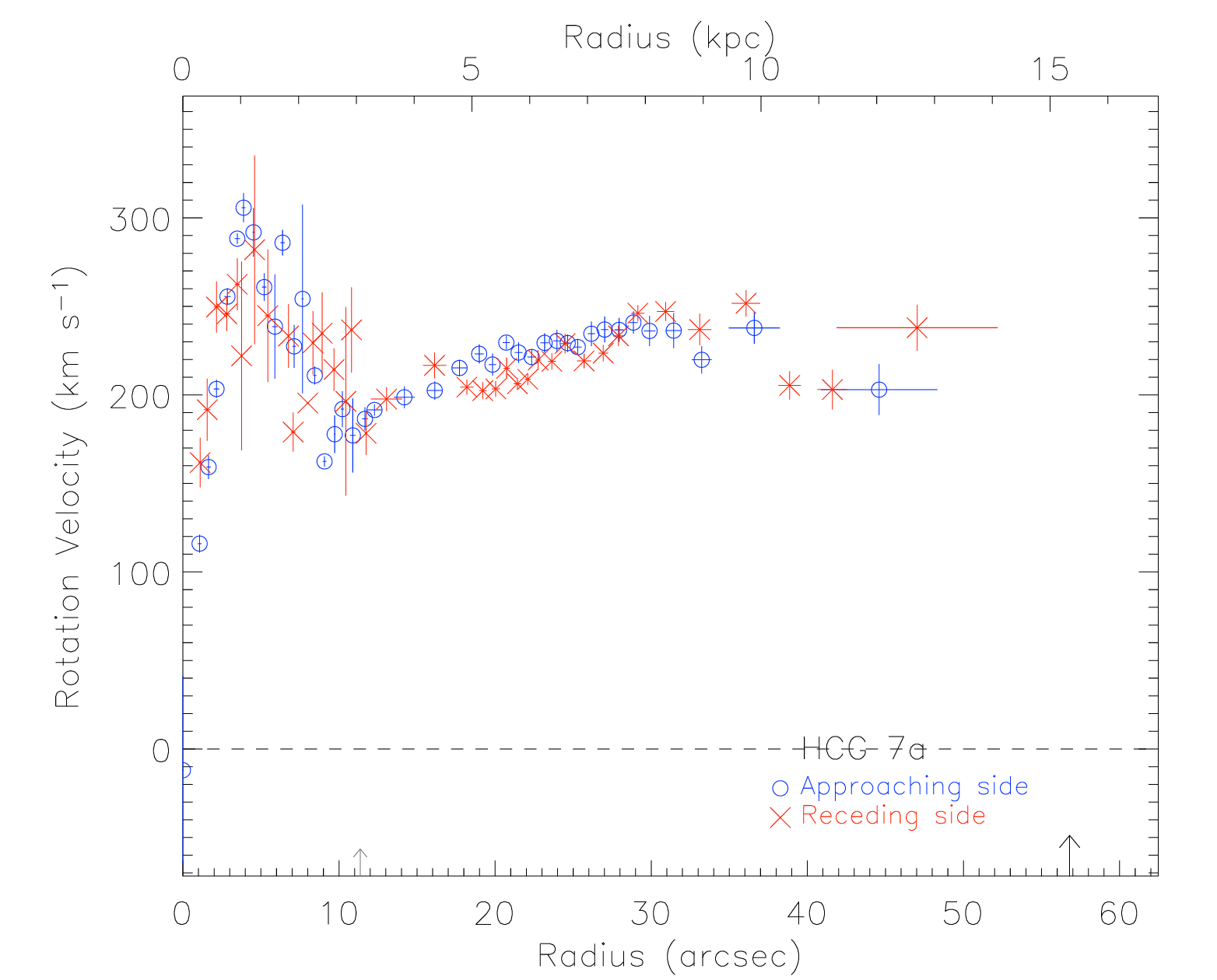}
\end{figure*}

\begin{figure*}[h!]
\includegraphics[width=\columnwidth]{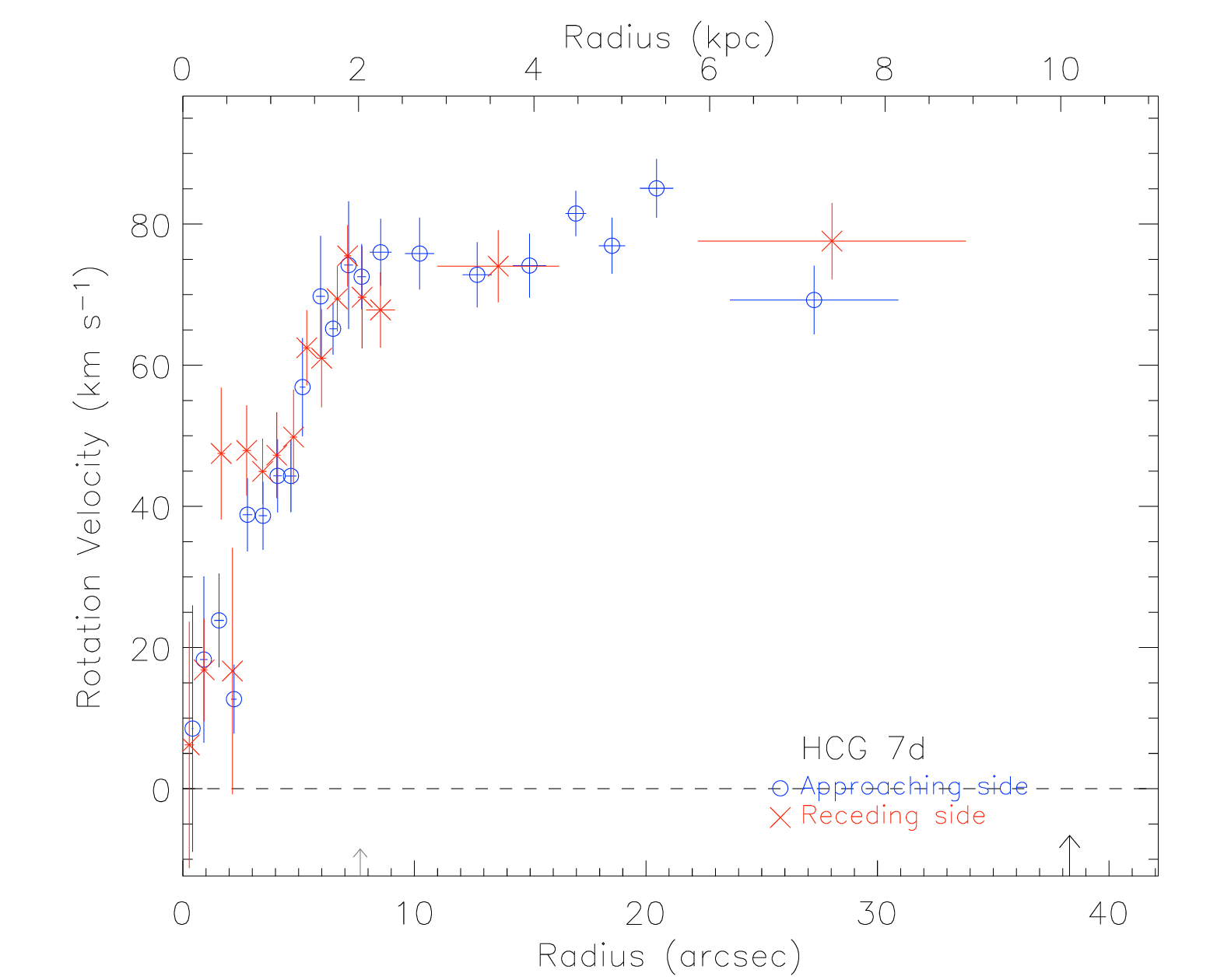}
\includegraphics[width=\columnwidth]{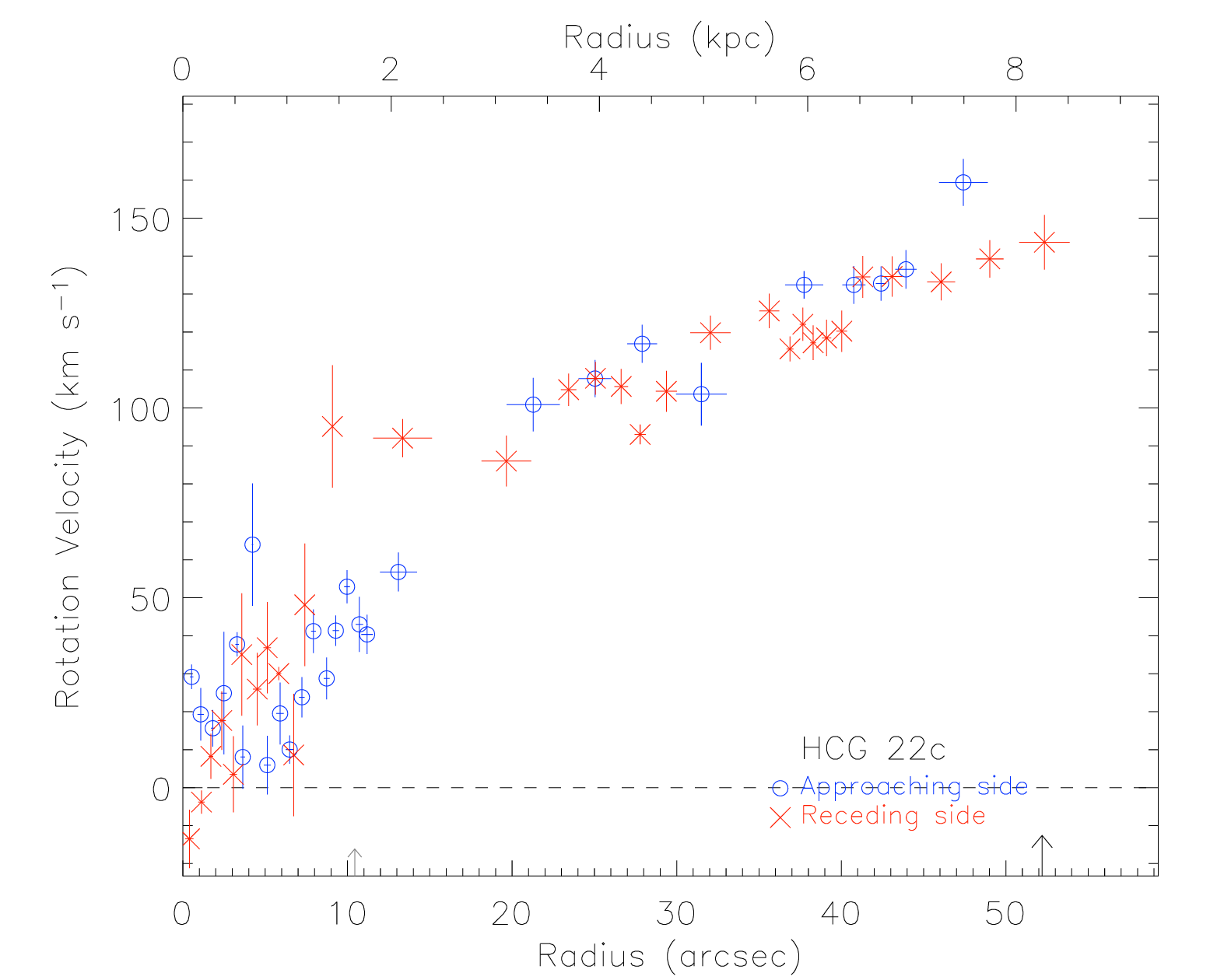} 
\caption{Rotation curves in this sample. The PA was determinated automatically from the model. The center and inclination were fixed (mophological center and photometic inclination, respectively). The black vertical arrow in the x-axis represents the radius R$_{25}$ while the smaller grey arrow in the x-axis represents the transition radius that is defined by the first ring that contains more than 25 uncorrelated bins in the velocity field (see Epinat et al. 2008a).}
\label{rot_cur1}
\end{figure*}
\clearpage

\begin{figure*}[h!]
\includegraphics[width=\columnwidth]{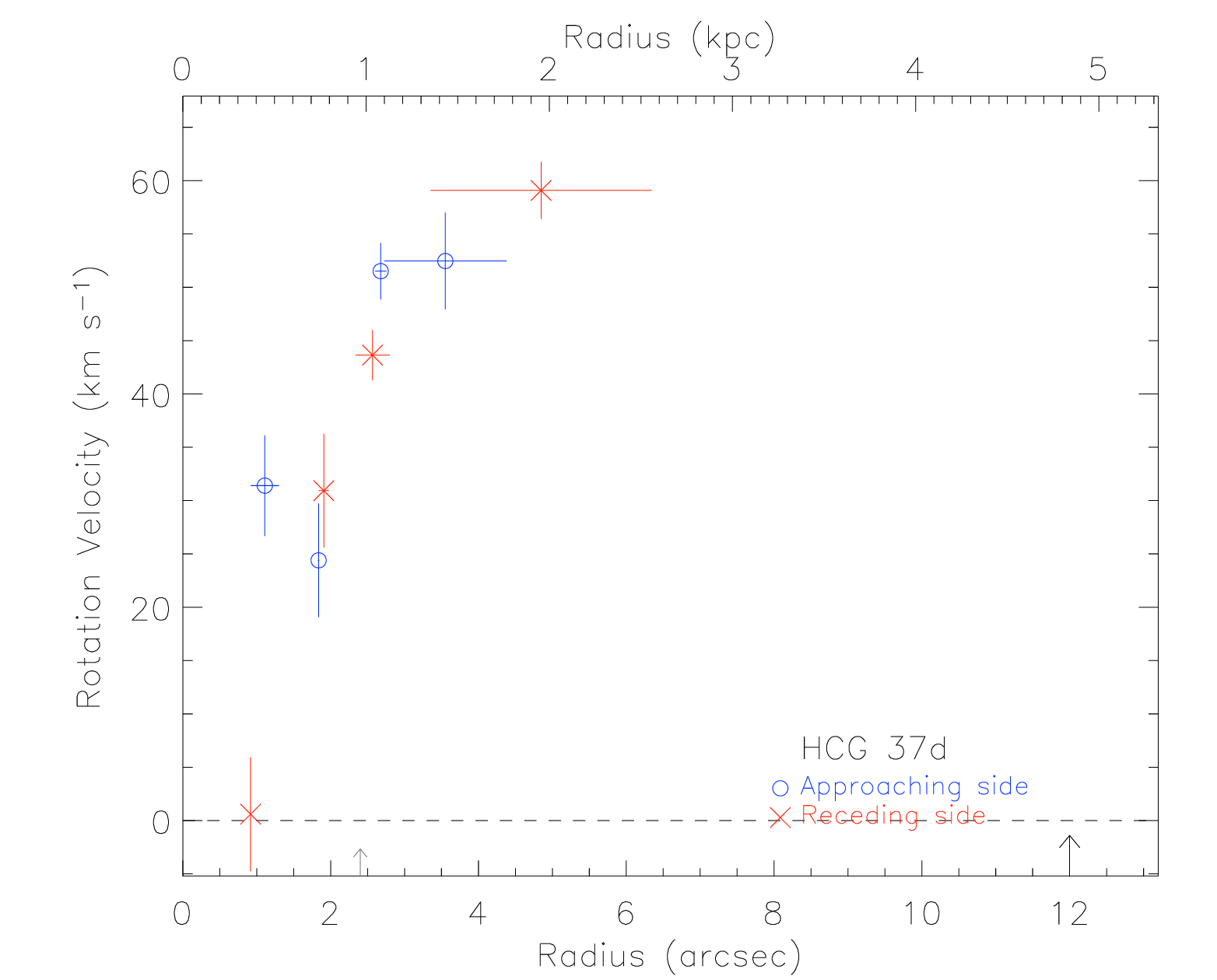}
\includegraphics[width=\columnwidth]{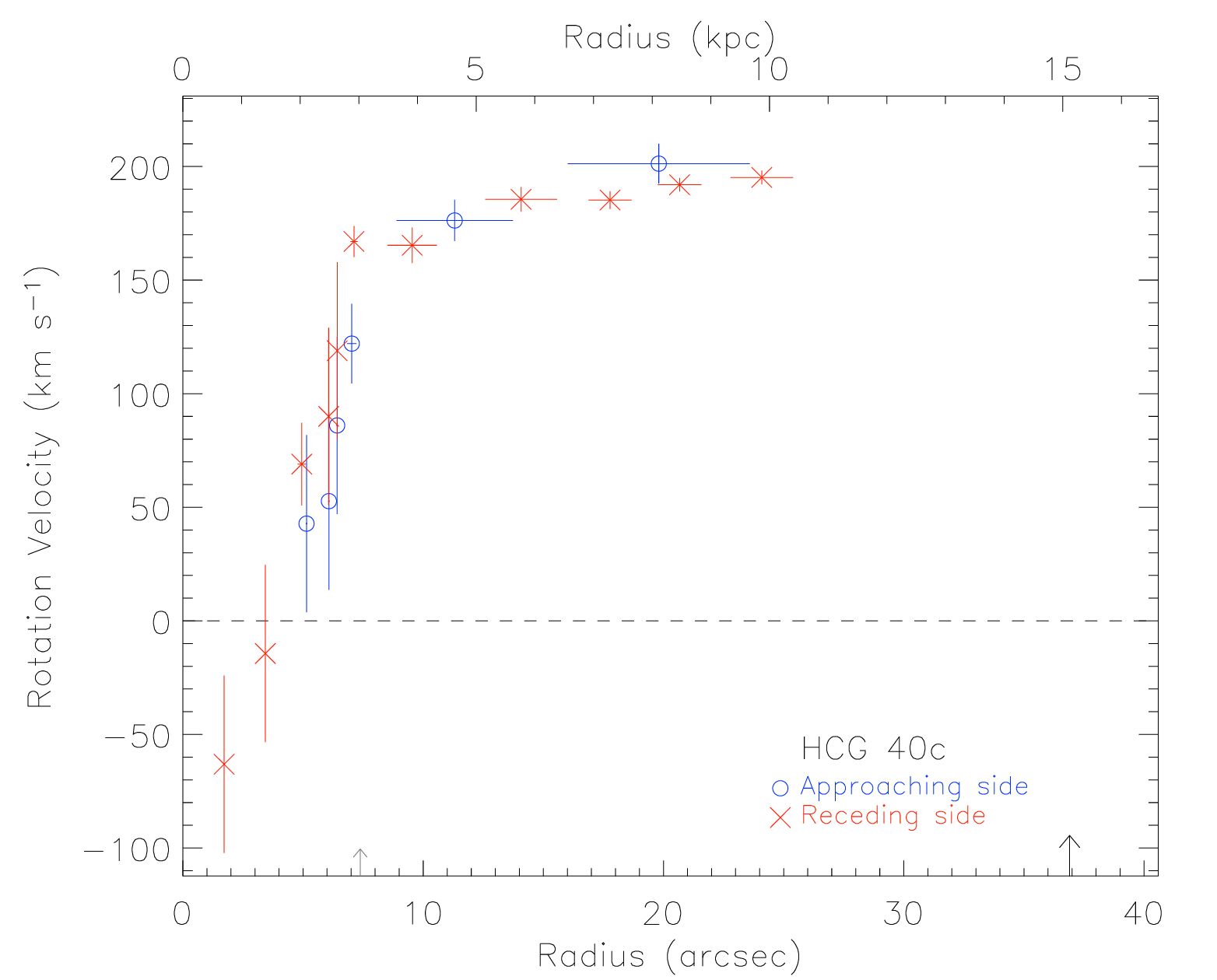}
\end{figure*}

\begin{figure*}[h!]
\includegraphics[width=\columnwidth]{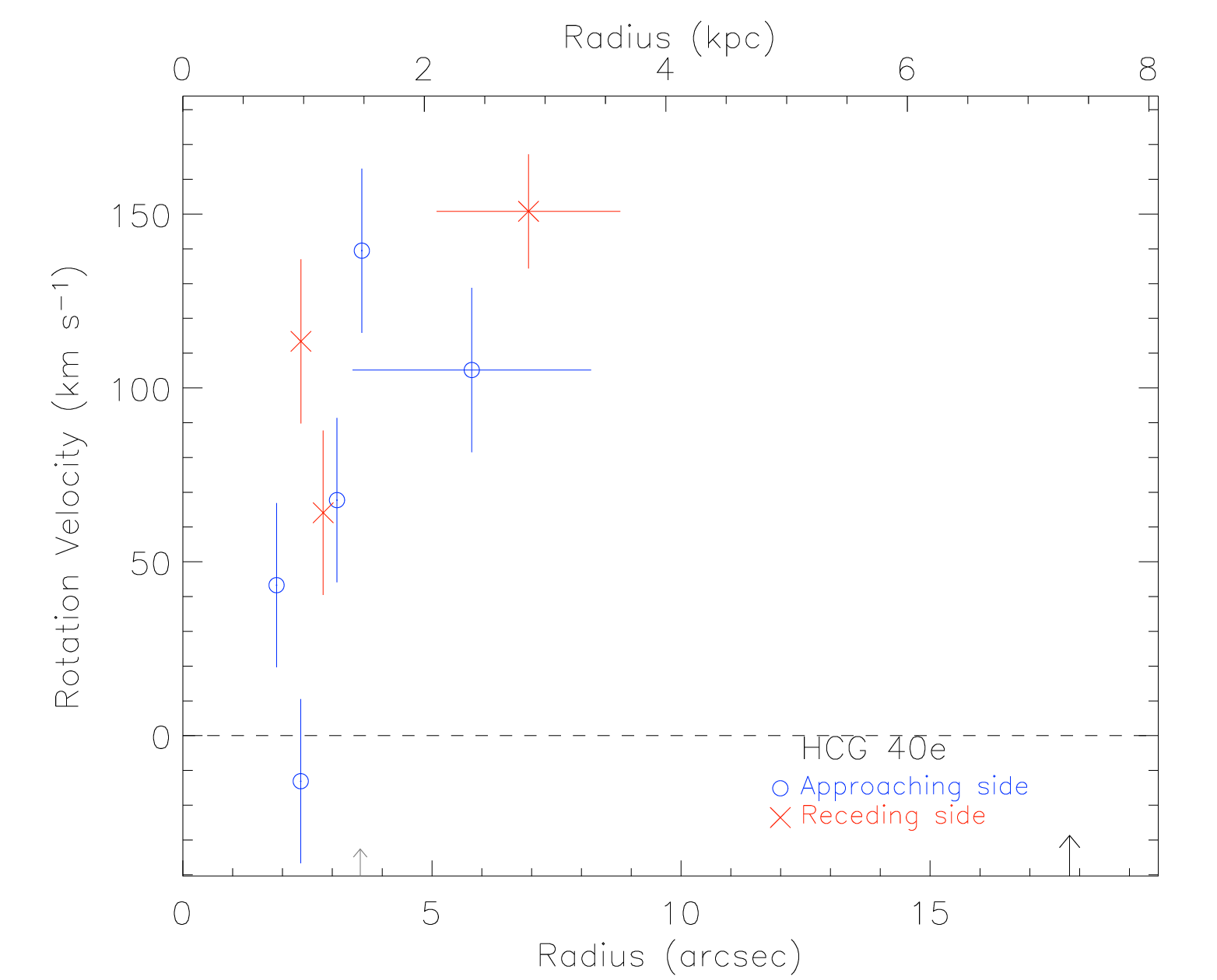}
\includegraphics[width=\columnwidth]{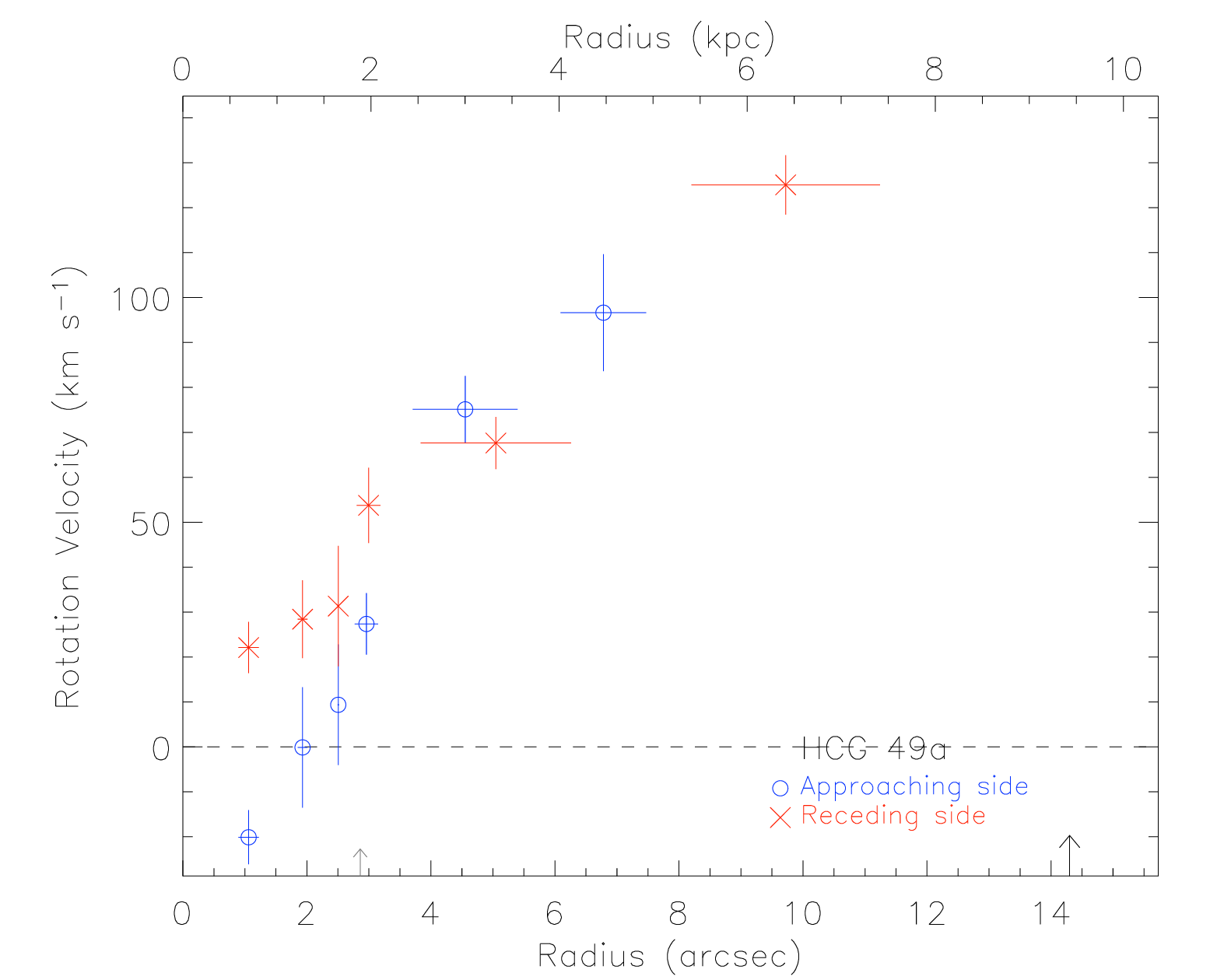} 
\end{figure*}

\begin{figure*}[h!]
\includegraphics[width=\columnwidth]{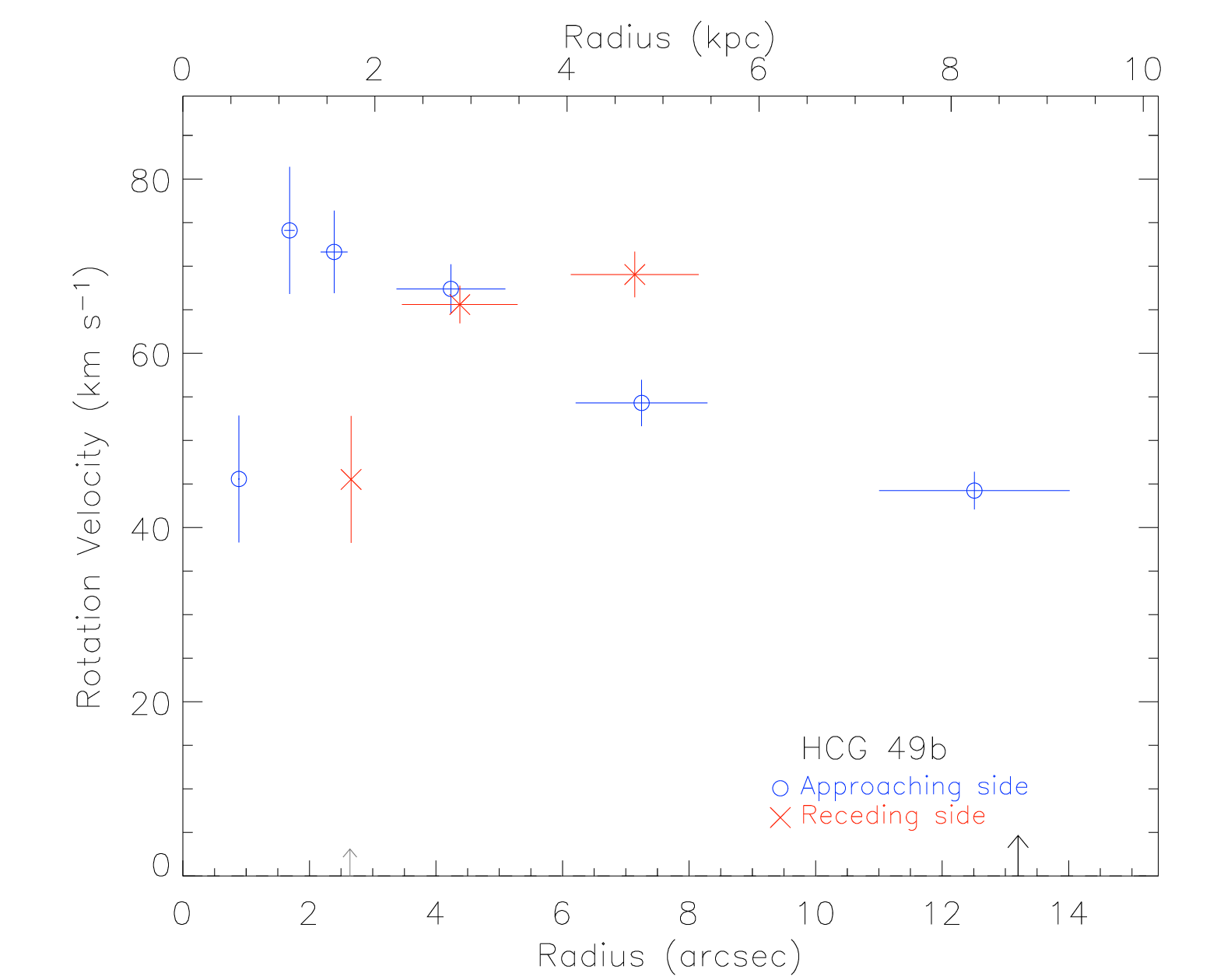}
\includegraphics[width=\columnwidth]{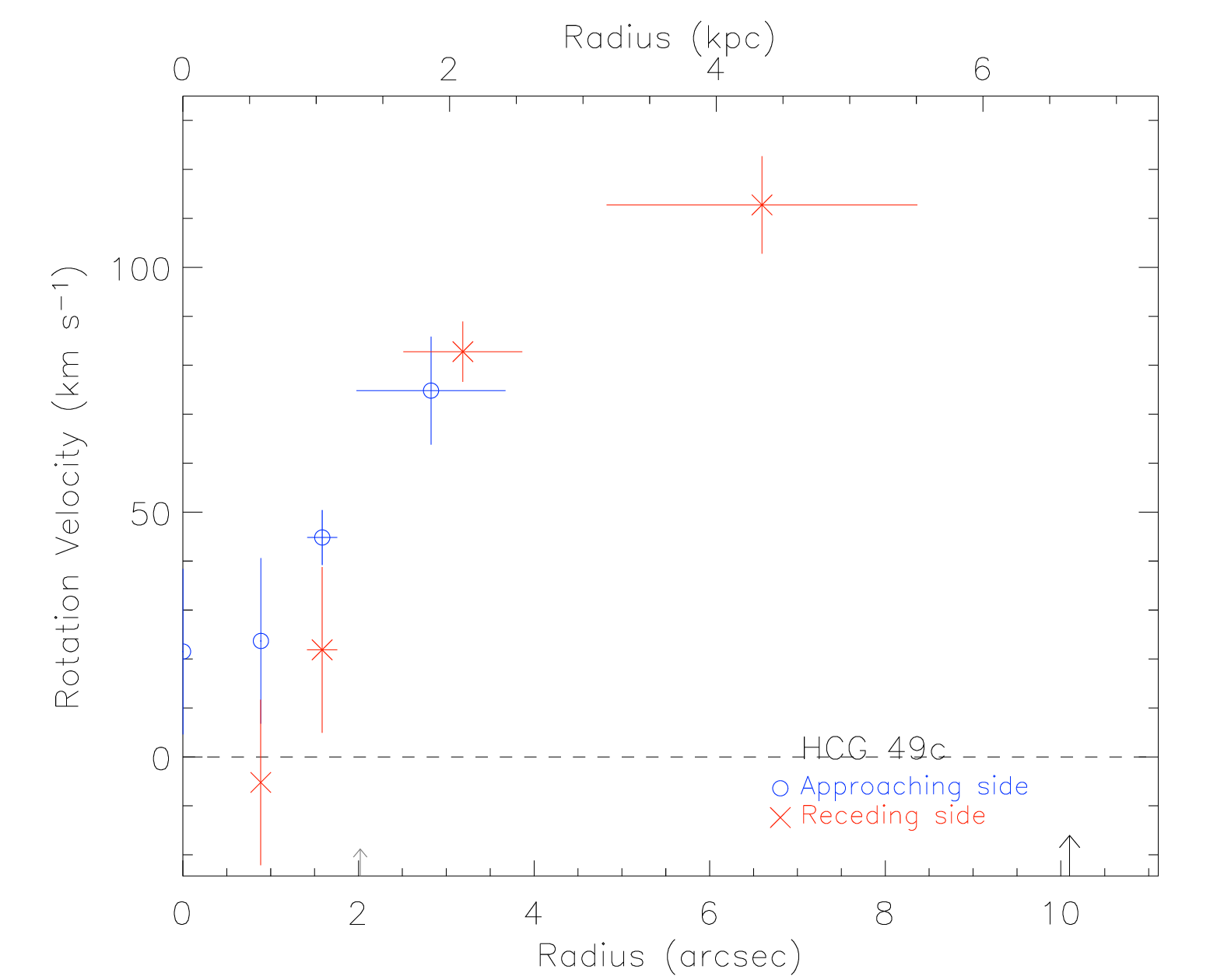}
\caption{See caption of Fig. \ref{rot_cur1}}
\label{rot_cur2}
\end{figure*}
\clearpage

\begin{figure*}[h!]
\includegraphics[width=\columnwidth]{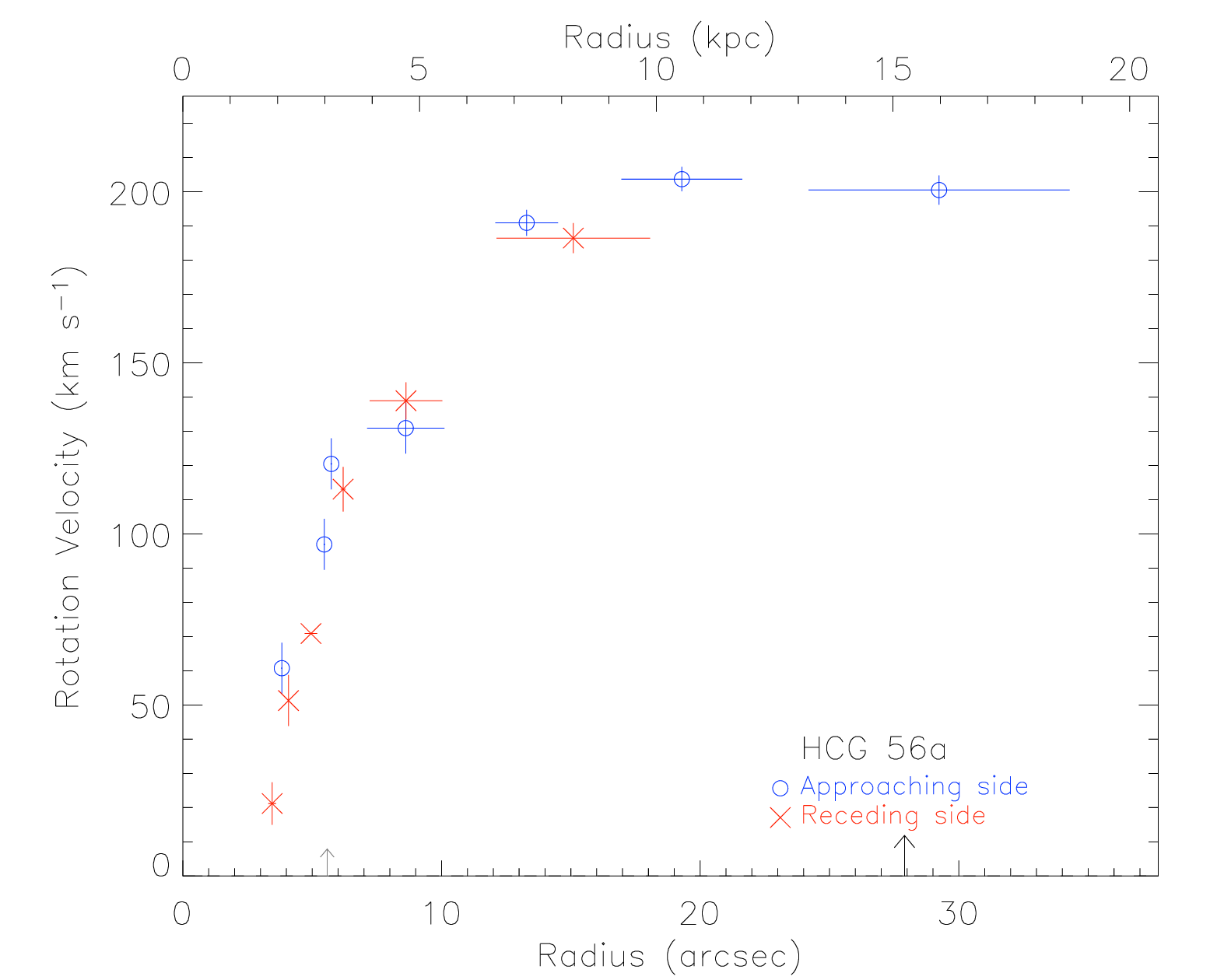}
\includegraphics[width=\columnwidth]{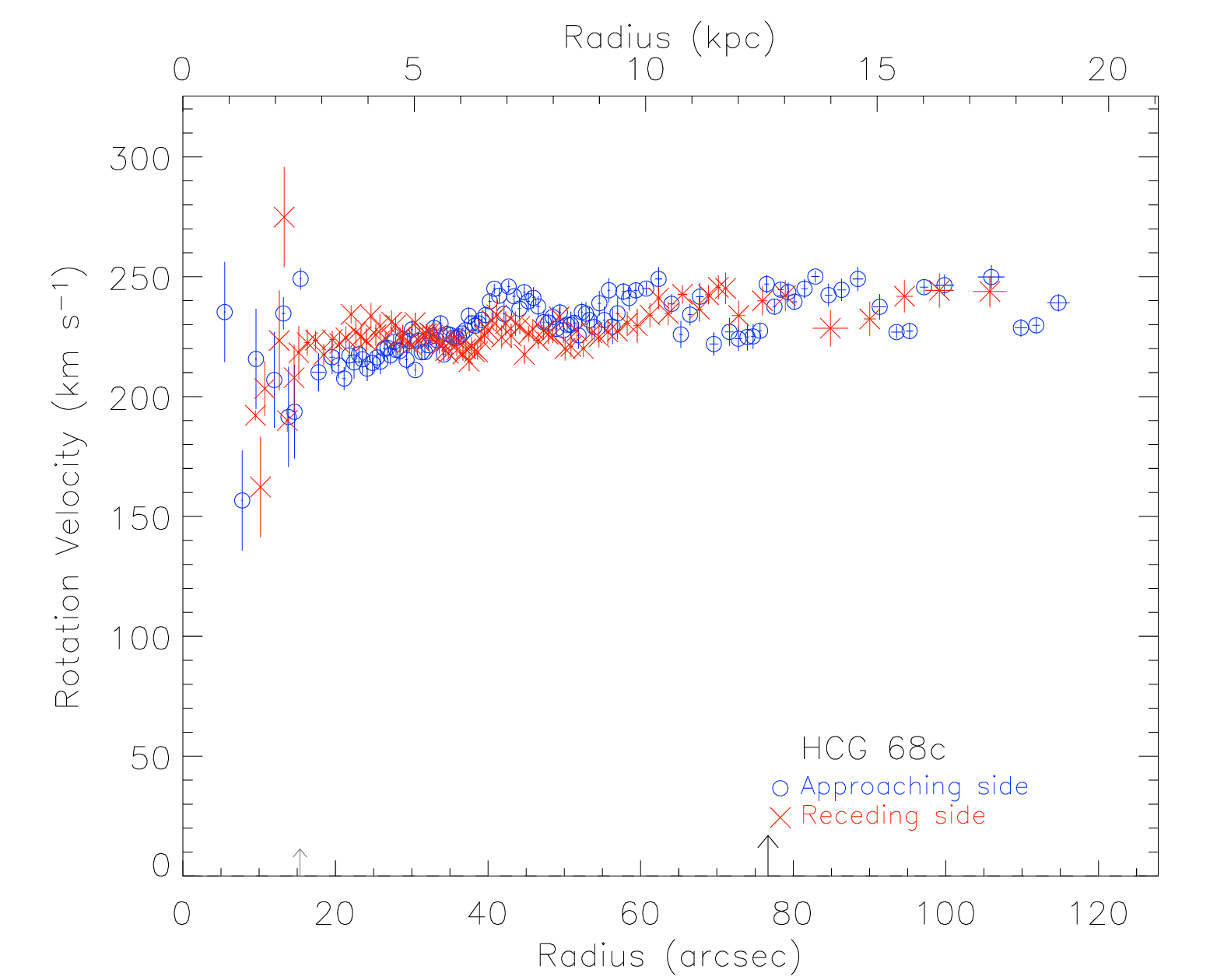} 
\end{figure*}

\begin{figure*}[h!]
\includegraphics[width=\columnwidth]{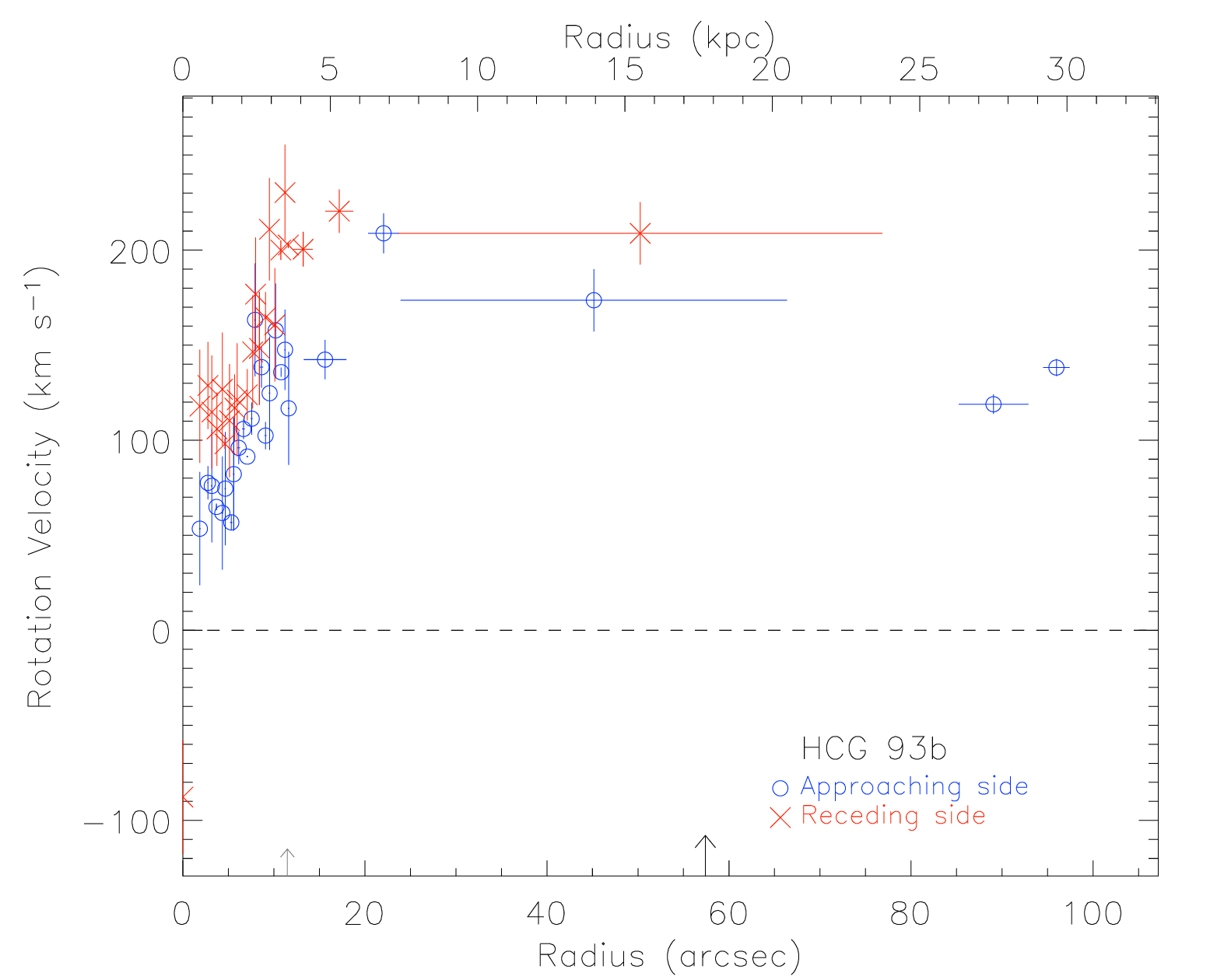}
\includegraphics[width=\columnwidth]{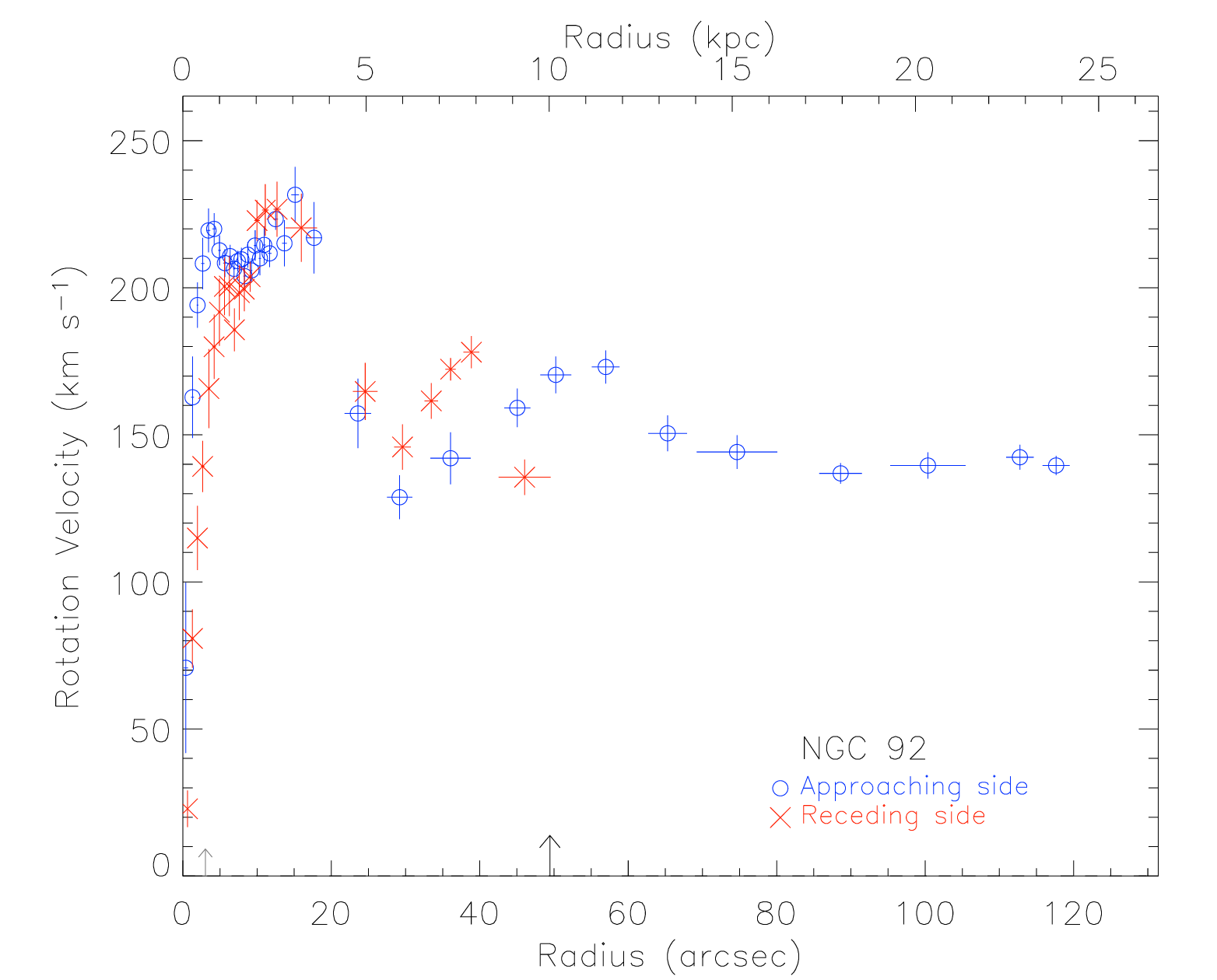} 
\caption{See caption of Fig. \ref{rot_cur1}}
\label{rot_cur3}
\end{figure*}

\end{appendix}

\clearpage

\Online

\begin{appendix} 

\section{Rotation Curve Tables (long tables are placed at the end of this appendix)}

\begin{table}
\caption{HCG 2b rotation curve}
\begin{tabular}{cccccccc}
\noalign{\medskip} \hline
           r  &  $\sigma_r$  &             r   &    $\sigma_r$  &          v   & $\sigma_v$   &            N   &              side \\
(kpc) &(kpc) &(\arcsec) & (\arcsec)& (km/s)& (km/s)&bins \\ 
(1)&(2)&(3)&(4)&(5)&(6)&(7)&(8)\\
\hline
            0.12  &            0.07  &             0.4  &             0.2  &              49  &               3  &               4  &               a \\
            0.15  &            0.03  &             0.5  &             0.1  &             -26  &               1  &               3  &               r \\
            0.32  &            0.05  &             1.1  &             0.2  &              12  &               7  &               9  &               r \\
            0.32  &            0.05  &             1.1  &             0.2  &              50  &               5  &               9  &               a \\
            0.52  &            0.05  &             1.8  &             0.2  &              55  &               5  &              14  &               r \\
\end{tabular}
\\(1), (3) Galactic radius. (2), (4) Dispersion around the galactic radius. (5) Rotation velocity. (6) Dispersion on the rotation velocity. (7) Number of velocity bins. (8) Receding -- r -- or approaching -- a -- side.
\end{table}

\begin{table}
\caption{HCG 7d rotation curve}
\begin{tabular}{cccccccc}
\noalign{\medskip} \hline
           r   &  $\sigma_r$  &             r   &    $\sigma_r$   &          v   & $\sigma_v$  &            N   &              side \\
(kpc) &(kpc) &(\arcsec) & (\arcsec)& (km/s)& (km/s) & bins\\ 
(1)&(2)&(3)&(4)&(5)&(6)&(7)&(8)\\
\hline
            0.07  &            0.05  &             0.3  &             0.2  &               6  &              17  &               2  &               r \\
            0.11  &            0.00  &             0.4  &             0.0  &               8  &              17  &               1  &               a \\
            0.24  &            0.04  &             0.9  &             0.2  &              18  &              11  &               6  &               a \\
            0.24  &            0.04  &             0.9  &             0.1  &              16  &               7  &               4  &               r \\
            0.41  &            0.05  &             1.6  &             0.2  &              23  &               6  &               9  &               a \\
\end{tabular}
\\(1), (3) Galactic radius. (2), (4) Dispersion around the galactic radius. (5) Rotation velocity. (6) Dispersion on the rotation velocity. (7) Number of velocity bins. (8) Receding -- r -- or approaching -- a -- side.
\end{table}

\begin{table}
\caption{HCG 22c rotation curve}
\begin{tabular}{cccccccc}
\noalign{\medskip} \hline
           r   &  $\sigma_r$  &             r   &    $\sigma_r$   &          v   & $\sigma_v$  &            N   &              side \\
(kpc) &(kpc) &(\arcsec) & (\arcsec)& (km/s)& (km/s) & bins\\
(1)&(2)&(3)&(4)&(5)&(6)&(7)&(8)\\
\hline
            0.06  &            0.04  &             0.4  &             0.2  &             -13  &               7  &               4  &               r \\
            0.08  &            0.02  &             0.5  &             0.1  &              29  &               3  &               3  &               a \\
            0.17  &            0.03  &             1.1  &             0.2  &              19  &               6  &               7  &               a \\
            0.18  &            0.03  &             1.1  &             0.2  &              -3  &               3  &               9  &               r \\
            0.27  &            0.02  &             1.7  &             0.1  &               8  &               5  &              10  &               r \\
\end{tabular}
\\(1), (3) Galactic radius. (2), (4) Dispersion around the galactic radius. (5) Rotation velocity. (6) Dispersion on the rotation velocity. (7) Number of velocity bins. (8) Receding -- r -- or approaching -- a -- side.
\end{table}

\begin{table}
\caption{HCG 37d Rotation Curve }
\begin{tabular}{cccccccc}
\noalign{\medskip} \hline
           r   &  $\sigma_r$   &             r   &    $\sigma_r$   &          v   & $\sigma_v$   &            N  &              side \\
(kpc) & (kpc) &(\arcsec) &(\arcsec) & (km/s)& (km/s) & bins\\ 
(1)&(2)&(3)&(4)&(5)&(6)&(7)&(8)\\
\hline
            0.37  &            0.00  &             0.9  &             0.0  &               0  &               5  &               1  &               r \\
            0.45  &            0.08  &             1.1  &             0.2  &              31  &               4  &               2  &               a \\
            0.74  &            0.00  &             1.8  &             0.0  &              24  &               5  &               1  &               a \\
            0.77  &            0.03  &             1.9  &             0.1  &              30  &               5  &               2  &               r \\
            1.04  &            0.09  &             2.6  &             0.2  &              43  &               2  &               2  &               r \\
            1.08  &            0.03  &             2.7  &             0.1  &              51  &               2  &               2  &               a \\
            1.43  &            0.33  &             3.6  &             0.8  &              52  &               4  &               7  &               a \\
            1.96  &            0.60  &             4.8  &             1.5  &              59  &               2  &              19  &               r \\
\hline
\label{rclabel_hcg37d}
\end{tabular}
\\(1), (3) Galactic radius. (2), (4) Dispersion around the galactic radius. (5) Rotation velocity. (6) Dispersion on the rotation velocity. (7) Number of velocity bins. (8) Receding -- r -- or approaching -- a -- side.
\end{table}

\begin{table}
\caption{HCG 40c rotation curve}
\begin{tabular}{cccccccc}
\noalign{\medskip} \hline
           r   &  $\sigma_r$   &             r   &    $\sigma_r$   &          v   & $\sigma_v$   &            N  &              side \\
(kpc) & (kpc) &(\arcsec) &(\arcsec) & (km/s)& (km/s) & bins\\
(1)&(2)&(3)&(4)&(5)&(6)&(7)&(8)\\
\hline
            0.70  &            0.00  &             1.7  &             0.0  &             -63  &              38  &               1  &               r \\
            1.41  &            0.00  &             3.4  &             0.0  &             -14  &              38  &               1  &               r \\
            2.03  &            0.07  &             4.9  &             0.2  &              68  &              18  &               2  &               r \\
            2.11  &            0.00  &             5.1  &             0.0  &              42  &              38  &               1  &               a \\
            2.49  &            0.00  &             6.1  &             0.0  &              90  &              38  &               2  &               r \\
\end{tabular}
\\(1), (3) Galactic radius. (2), (4) Dispersion around the galactic radius. (5) Rotation velocity. (6) Dispersion on the rotation velocity. (7) Number of velocity bins. (8) Receding -- r -- or approaching -- a -- side.
\end{table}

\begin{table}
\caption{HCG 40e rotation curve}
\begin{tabular}{cccccccc}
\noalign{\medskip} \hline
           r   &  $\sigma_r$   &             r   &    $\sigma_r$   &          v   & $\sigma_v$   &            N  &              side \\
(kpc) & (kpc) &(\arcsec) &(\arcsec) & (km/s)& (km/s) & bins\\
(1)&(2)&(3)&(4)&(5)&(6)&(7)&(8)\\
\hline
            0.78  &            0.00  &             1.9  &             0.0  &              43  &              23  &               1  &               a \\
            0.98  &            0.00  &             2.4  &             0.0  &             -13  &              23  &               1  &               a \\
            0.98  &            0.00  &             2.4  &             0.0  &             113  &              23  &               1  &               r \\
            1.16  &            0.00  &             2.8  &             0.0  &              64  &              23  &               1  &               r \\
            1.28  &            0.00  &             3.1  &             0.0  &              67  &              23  &               1  &               a \\
            1.48  &            0.00  &             3.6  &             0.0  &             139  &              23  &               1  &               a \\
            2.39  &            0.99  &             5.8  &             2.4  &             105  &              23  &               3  &               a \\
            2.86  &            0.76  &             6.9  &             1.8  &             150  &              16  &               3  &               r \\
\hline
\label{rclabel_hcg40e}
\end{tabular}
\\(1), (3) Galactic radius. (2), (4) Dispersion around the galactic radius. (5) Rotation velocity. (6) Dispersion on the rotation velocity. (7) Number of velocity bins. (8) Receding -- r -- or approaching -- a -- side.
\end{table}

\begin{table}
\caption{HCG 49a rotation curve}
\begin{tabular}{cccccccc}
\noalign{\medskip} \hline
           r   &  $\sigma_r$   &             r   &    $\sigma_r$   &          v   & $\sigma_v$   &            N  &              side \\
(kpc) & (kpc) &(\arcsec) &(\arcsec) & (km/s)& (km/s) & bins\\
(1)&(2)&(3)&(4)&(5)&(6)&(7)&(8)\\
\hline
            0.70  &            0.11  &             1.1  &             0.2  &             -20  &               6  &               3  &               a \\
            0.70  &            0.11  &             1.1  &             0.2  &              22  &               5  &               3  &               r \\
            1.27  &            0.05  &             1.9  &             0.1  &               0  &              13  &               4  &               a \\
            1.27  &            0.05  &             1.9  &             0.1  &              28  &               8  &               4  &               r \\
            1.65  &            0.00  &             2.5  &             0.0  &              31  &              13  &               1  &               r \\
\end{tabular}
\\(1), (3) Galactic radius. (2), (4) Dispersion around the galactic radius. (5) Rotation velocity. (6) Dispersion on the rotation velocity. (7) Number of velocity bins. (8) Receding -- r -- or approaching -- a -- side.
\end{table}

\begin{table}
\caption{HCG 49b rotation curve}
\begin{tabular}{cccccccc}
\noalign{\medskip} \hline
           r   &  $\sigma_r$   &             r   &    $\sigma_r$   &          v   & $\sigma_v$   &            N  &              side \\
(kpc) & (kpc) &(\arcsec) &(\arcsec) & (km/s)& (km/s) & bins\\
(1)&(2)&(3)&(4)&(5)&(6)&(7)&(8)\\
\hline
            0.58  &            0.00  &             0.9  &             0.0  &              45  &               7  &               1  &               a \\
            1.11  &            0.05  &             1.7  &             0.1  &              74  &               7  &               3  &               a \\
            1.57  &            0.14  &             2.4  &             0.2  &              71  &               4  &               3  &               a \\
            1.75  &            0.00  &             2.7  &             0.0  &              45  &               7  &               1  &               r \\
            2.79  &            0.57  &             4.2  &             0.9  &              67  &               2  &              24  &               a \\
            2.88  &            0.60  &             4.4  &             0.9  &              65  &               2  &              24  &               r \\
            4.71  &            0.67  &             7.1  &             1.0  &              69  &               2  &              19  &               r \\
            4.78  &            0.69  &             7.3  &             1.0  &              54  &               2  &              24  &               a \\
            8.24  &            0.99  &            12.5  &             1.5  &              44  &               2  &              24  &               a \\
\hline
\label{rclabel_hcg49b}
\end{tabular}
\\(1), (3) Galactic radius. (2), (4) Dispersion around the galactic radius. (5) Rotation velocity. (6) Dispersion on the rotation velocity. (7) Number of velocity bins. (8) Receding -- r -- or approaching -- a -- side.
\end{table}

\begin{table}
\caption{HCG 49c rotation curve}
\begin{tabular}{cccccccc}
\noalign{\medskip} \hline
           r   &  $\sigma_r$   &             r   &    $\sigma_r$   &          v   & $\sigma_v$   &            N  &              side \\
(kpc) & (kpc) &(\arcsec) &(\arcsec) & (km/s)& (km/s) & bins\\
(1)&(2)&(3)&(4)&(5)&(6)&(7)&(8)\\
\hline
            0.00  &            0.00  &             0.0  &             0.0  &              21  &              16  &               1  &               a \\
            0.58  &            0.00  &             0.9  &             0.0  &              -5  &              16  &               1  &               r \\
            0.59  &            0.00  &             0.9  &             0.0  &              23  &              16  &               1  &               a \\
            1.04  &            0.11  &             1.6  &             0.2  &              21  &              16  &               3  &               r \\
            1.05  &            0.11  &             1.6  &             0.2  &              44  &               5  &               3  &               a \\
            1.86  &            0.56  &             2.8  &             0.8  &              74  &              11  &               7  &               a \\
            2.10  &            0.45  &             3.2  &             0.7  &              82  &               6  &              16  &               r \\
            4.34  &            1.16  &             6.6  &             1.8  &             112  &               9  &              10  &               r \\
\hline
\label{rclabel_hcg49c}
\end{tabular}
\\(1), (3) Galactic radius. (2), (4) Dispersion around the galactic radius. (5) Rotation velocity. (6) Dispersion on the rotation velocity. (7) Number of velocity bins. (8) Receding -- r -- or approaching -- a -- side.
\end{table}

\begin{table}
\caption{HCG 56a rotation curve}
\begin{tabular}{cccccccc}
\noalign{\medskip} \hline
           r   &  $\sigma_r$   &             r   &    $\sigma_r$   &          v   & $\sigma_v$   &            N  &              side \\
(kpc) & (kpc) &(\arcsec) &(\arcsec) & (km/s)& (km/s) & bins\\
(1)&(2)&(3)&(4)&(5)&(6)&(7)&(8)\\
\hline
            1.89  &            0.08  &             3.5  &             0.1  &              21  &               6  &               2  &               r \\
            2.09  &            0.00  &             3.8  &             0.0  &              60  &               7  &               1  &               a \\
            2.23  &            0.00  &             4.1  &             0.0  &              51  &               7  &               1  &               r \\
            2.71  &            0.13  &             5.0  &             0.2  &              70  &               0  &               2  &               r \\
            2.99  &            0.00  &             5.5  &             0.0  &              96  &               7  &               1  &               a \\
\end{tabular}
\\(1), (3) Galactic radius. (2), (4) Dispersion around the galactic radius. (5) Rotation velocity. (6) Dispersion on the rotation velocity. (7) Number of velocity bins. (8) Receding -- r -- or approaching -- a -- side.
\end{table}

\begin{table}
\caption{HCG 79d rotation curve}
\begin{tabular}{cccccccc}
\noalign{\medskip} \hline
           r   &  $\sigma_r$   &             r   &    $\sigma_r$   &          v   & $\sigma_v$   &            N  &              side \\
(kpc) & (kpc) &(\arcsec) &(\arcsec) & (km/s)& (km/s) & bins\\
(1)&(2)&(3)&(4)&(5)&(6)&(7)&(8)\\
\hline
            0.00  &            0.00  &             0.0  &             0.0  &             -35  &              14  &               1  &               a \\
            0.33  &            0.00  &             1.1  &             0.0  &              38  &              14  &               1  &               r \\
            0.33  &            0.00  &             1.1  &             0.0  &               3  &              14  &               1  &               a \\
            0.51  &            0.00  &             1.7  &             0.0  &              26  &              14  &               1  &               a \\
            0.63  &            0.00  &             2.0  &             0.0  &              20  &              14  &               1  &               r \\
\end{tabular}
\\(1), (3) Galactic radius. (2), (4) Dispersion around the galactic radius. (5) Rotation velocity. (6) Dispersion on the rotation velocity. (7) Number of velocity bins. (8) Receding -- r -- or approaching -- a -- side.
\end{table}

\begin{table}
\caption{HCG 93b rotation curve}
\begin{tabular}{cccccccc}
\noalign{\medskip} \hline
           r   &  $\sigma_r$   &             r   &    $\sigma_r$   &          v   & $\sigma_v$   &            N  &              side \\
(kpc) & (kpc) &(\arcsec) &(\arcsec) & (km/s)& (km/s) & bins\\
(1)&(2)&(3)&(4)&(5)&(6)&(7)&(8)\\
\hline
            0.00  &            0.00  &             0.0  &             0.0  &             -87  &              29  &               1  &               r \\
            0.58  &            0.00  &             1.9  &             0.0  &              53  &              29  &               1  &               a \\
            0.58  &            0.00  &             1.9  &             0.0  &             117  &              29  &               1  &               r \\
            0.85  &            0.02  &             2.7  &             0.1  &              77  &               8  &               2  &               a \\
            0.85  &            0.02  &             2.7  &             0.1  &             128  &              22  &               2  &               r \\
\end{tabular}
\\(1), (3) Galactic radius. (2), (4) Dispersion around the galactic radius. (5) Rotation velocity. (6) Dispersion on the rotation velocity. (7) Number of velocity bins. (8) Receding -- r -- or approaching -- a -- side.
\end{table}

\begin{table}
\caption{NGC 92 rotation curve}
\begin{tabular}{cccccccc}
\noalign{\medskip} \hline
           r   &  $\sigma_r$   &             r   &    $\sigma_r$   &          v   & $\sigma_v$   &            N  &              side \\
(kpc) & (kpc) &(\arcsec) &(\arcsec) & (km/s)& (km/s) & bins\\
(1)&(2)&(3)&(4)&(5)&(6)&(7)&(8)\\
\hline
            0.08  &            0.06  &             0.4  &             0.3  &              70  &              28  &               3  &               a \\
            0.13  &            0.00  &             0.6  &             0.0  &              22  &               6  &               2  &               r \\
            0.26  &            0.04  &             1.3  &             0.2  &              80  &               9  &               6  &               r \\
            0.26  &            0.04  &             1.3  &             0.2  &             162  &              13  &               6  &               a \\
            0.40  &            0.02  &             2.0  &             0.1  &             114  &              10  &               6  &               r \\
\end{tabular}
\\(1), (3) Galactic radius. (2), (4) Dispersion around the galactic radius. (5) Rotation velocity. (6) Dispersion on the rotation velocity. (7) Number of velocity bins. (8) Receding -- r -- or approaching -- a -- side.
\end{table}

\longtab{1}{
\begin{longtable}{cccccccc}
\caption{HCG 2a rotation curve}\\
\hline
r   &  $\sigma_r$   &             r   &    $\sigma_r$   &          v   & $\sigma_v$   &            N  &              side \\
(kpc) & (kpc) &(\arcsec) &(\arcsec) & (km/s)& (km/s) & bins\\
(1)&(2)&(3)&(4)&(5)&(6)&(7)&(8)\\
\hline
\endfirsthead
\caption{continued.}\\
\hline
r   &  $\sigma_r$   &             r   &    $\sigma_r$   &          v   & $\sigma_v$   &            N  &              side \\
(kpc) & (kpc) &(\arcsec) &(\arcsec) & (km/s)& (km/s) & bins\\
(1)&(2)&(3)&(4)&(5)&(6)&(7)&(8)\\
\hline
\endhead
\hline
\endfoot
            0.18  &            0.01  &             0.6  &             0.0  &             -49  &              13  &               3  &               a \\
            0.18  &            0.01  &             0.6  &             0.0  &              62  &              16  &               3  &               r \\
            0.33  &            0.03  &             1.2  &             0.1  &              60  &              10  &               5  &               r \\
            0.33  &            0.03  &             1.2  &             0.1  &             -32  &               9  &               5  &               a \\
            0.52  &            0.05  &             1.8  &             0.2  &              55  &               5  &               7  &               r \\
            0.52  &            0.06  &             1.8  &             0.2  &             -24  &               4  &               9  &               a \\
            0.70  &            0.06  &             2.5  &             0.2  &              45  &               3  &              11  &               r \\
            0.71  &            0.06  &             2.5  &             0.2  &             -20  &               3  &              12  &               a \\
\end{longtable}
}

\longtab{2}{
\begin{longtable}{cccccccc}
\caption{HCG 2c rotation curve}\\
\hline
r   &  $\sigma_r$   &             r   &    $\sigma_r$   &          v   & $\sigma_v$   &            N  &              side \\
(kpc) & (kpc) &(\arcsec) &(\arcsec) & (km/s)& (km/s) & bins\\
(1)&(2)&(3)&(4)&(5)&(6)&(7)&(8)\\
\hline
\endfirsthead
\caption{continued.}\\
\hline
r   &  $\sigma_r$   &             r   &    $\sigma_r$   &          v   & $\sigma_v$   &            N  &              side \\
(kpc) & (kpc) &(\arcsec) &(\arcsec) & (km/s)& (km/s) & bins\\
(1)&(2)&(3)&(4)&(5)&(6)&(7)&(8)\\
\hline
\endhead
\hline
\endfoot
            0.07  &            0.06  &             0.3  &             0.2  &              32  &               4  &               2  &               r \\
            0.12  &            0.00  &             0.4  &             0.0  &               0  &              25  &               1  &               a \\
            0.23  &            0.02  &             0.8  &             0.1  &             -11  &              12  &               2  &               a \\
            0.23  &            0.02  &             0.8  &             0.1  &              36  &               3  &               2  &               r \\
            0.42  &            0.05  &             1.5  &             0.2  &              36  &              13  &               3  &               r \\
            0.42  &            0.05  &             1.5  &             0.2  &             -30  &               7  &               6  &               a \\
            0.61  &            0.00  &             2.2  &             0.0  &              30  &              25  &               1  &               r \\
            0.67  &            0.03  &             2.4  &             0.1  &               4  &              25  &               3  &               a \\
\end{longtable}
}

\longtab{3}{
\begin{longtable}{cccccccc}
\caption{HCG 7a rotation curve}\\
\hline
r   &  $\sigma_r$   &             r   &    $\sigma_r$   &          v   & $\sigma_v$   &            N  &              side \\
(kpc) & (kpc) &(\arcsec) &(\arcsec) & (km/s)& (km/s) & bins\\
(1)&(2)&(3)&(4)&(5)&(6)&(7)&(8)\\
\hline
\endfirsthead
\caption{continued.}\\
\hline
r   &  $\sigma_r$   &             r   &    $\sigma_r$   &          v   & $\sigma_v$   &            N  &              side \\
(kpc) & (kpc) &(\arcsec) &(\arcsec) & (km/s)& (km/s) & bins\\
(1)&(2)&(3)&(4)&(5)&(6)&(7)&(8)\\
\hline
\endhead
\hline
\endfoot
            0.00  &            0.00  &             0.0  &             0.0  &             -12  &              53  &               1  &               a \\
            0.29  &            0.04  &             1.1  &             0.1  &             115  &               5  &               3  &               a \\
            0.30  &            0.03  &             1.1  &             0.1  &             161  &              14  &               4  &               r \\
            0.42  &            0.05  &             1.6  &             0.2  &             191  &              17  &               5  &               r \\
            0.44  &            0.04  &             1.6  &             0.1  &             159  &               6  &               4  &               a \\
            0.58  &            0.05  &             2.2  &             0.2  &             249  &              14  &              10  &               r \\
            0.58  &            0.05  &             2.2  &             0.2  &             203  &               5  &              10  &               a \\
            0.76  &            0.05  &             2.8  &             0.2  &             245  &               9  &               9  &               r \\
\end{longtable}
}

\longtab{4}{
\begin{longtable}{cccccccc}
\caption{HCG 68c rotation curve}\\
\hline
r   &  $\sigma_r$   &             r   &    $\sigma_r$   &          v   & $\sigma_v$   &            N  &              side \\
(kpc) & (kpc) &(\arcsec) &(\arcsec) & (km/s)& (km/s) & bins\\
(1)&(2)&(3)&(4)&(5)&(6)&(7)&(8)\\
\hline
\endfirsthead
\caption{continued.}\\
\hline
r   &  $\sigma_r$   &             r   &    $\sigma_r$   &          v   & $\sigma_v$   &            N  &              side \\
(kpc) & (kpc) &(\arcsec) &(\arcsec) & (km/s)& (km/s) & bins\\
(1)&(2)&(3)&(4)&(5)&(6)&(7)&(8)\\
\hline
\endhead
\hline
\endfoot
            0.90  &            0.00  &             5.5  &             0.0  &             235  &              20  &               1  &               a \\
            1.28  &            0.00  &             7.8  &             0.0  &             156  &              20  &               1  &               a \\
            1.57  &            0.00  &             9.5  &             0.0  &             192  &               2  &               2  &               r \\
            1.58  &            0.00  &             9.6  &             0.0  &             215  &              20  &               1  &               a \\
            1.68  &            0.00  &            10.2  &             0.0  &             162  &              20  &               1  &               r \\
            1.77  &            0.02  &            10.7  &             0.1  &             203  &              11  &               2  &               r \\
            1.97  &            0.00  &            12.0  &             0.0  &             206  &              19  &               2  &               a \\
            2.08  &            0.01  &            12.6  &             0.1  &             223  &              20  &               3  &               r \\
\end{longtable}
}

\end{appendix}

\end{document}